\documentclass{ptephy_v1}
\usepackage{graphics} %for dealing with figures.

\usepackage{bm,latexsym,amsmath,amssymb,amsfonts,mathrsfs,textcomp}
\usepackage[colorlinks=true, pdfstartview=FitV, linkcolor=red, citecolor=blue, urlcolor=blue]{hyperref}
\usepackage[normalem]{ulem}  % \sout{old text} for strikeout

\usepackage{adjustbox}
\usepackage{xcolor}

\newcommand{\tr}{\mathop{\mathrm{tr}}}

\newcommand{\diff}{\mathrm{d}} 
\newcommand{\rmi}{\mathrm{i}} 
\newcommand{\rme}{\mathrm{e}}

\newcommand{\para}{\parallel}

\newcommand{\tilalpha}{\widetilde{\alpha}}

\newcommand{\ptc}[1]{{\bar{#1}}}

\newcommand{\pNcal}{\ptc{\mathcal{N}}}

\newcommand\Lcal{\mathcal{L}}

\newcommand\Bcal{\mathcal{B}}

\newcommand\Ncal{\mathcal{N}}

\newcommand\diag{\operatorname{diag}}

\newcommand\SU{{\rm SU}}

\newcommand{\bra}[1]{\langle {#1} |}
\newcommand{\ket}[1]{| {#1} \rangle}

\newcommand{\braket}[1]{\langle {#1} \rangle}

\newcommand{\bx}{\bm{x}}

\newcommand{\bz}{\bm{z}}
\newcommand{\br}{\bm{r}}

\newcommand{\bp}{\bm{p}}

\newcommand{\bB}{\bm{B}}

\newcommand{\tilN}{\widetilde{N}}

\newcommand{\tild}{\widetilde{d}}

\newcommand{\U}{\text{U}}

\newcommand{\with}{\quad\mathrm{with}\quad}

\newcommand{\eff}{\mathrm{eff}}

\begin{document}

\title{
  One-pion exchange potential \\
  in a strong magnetic field
}

\author[1,a]{Daiki~Miura}
\affil[1]{Graduate School of Science and Technology, Niigata University,  Niigata, 950-2181, Japan}

\author[2,3,b]{Masaru~Hongo}
\affil[2]{Department of Physics, Niigata University, Niigata 950-2181, Japan}
\affil[3]{RIKEN Center for Interdisciplinary Theoretical and Mathematical Sciences (iTHEMS), RIKEN, Wako 351-0198, Japan}

\author[4,3,c]{Hidetoshi~Taya}
\affil[4]{Department of Physics, Keio University, 4-1-1 Hiyoshi, Kanagawa 223-8521, Japan}

\author[3,5,d]{Tetsuo~Hatsuda}
\affil[5]{Kavli Institute for the Physics and Mathematics of the Universe (Kavli IPMU), WPI, The University of Tokyo, Kashiwa, Chiba 277-8568, Japan}

\affil[a]{{\tt f23j004e@mail.cc.niigata-u.ac.jp}}
\affil[b]{{\tt hongo@phys.sc.niigata-u.ac.jp}}
\affil[c]{{\tt h\_taya@keio.jp}}
\affil[d]{{\tt thatsuda@riken.jp}}

\begin{abstract}
  We derive the one-pion exchange potential (OPEP) in the presence of a homogeneous magnetic field using chiral perturbation theory with nonrelativistic nucleons.
  Our approach is applicable not only to weak magnetic fields but also to strong ones up to {around} the pion-mass scale. The Green's function of charged pions is modified by the magnetic field, leading to changes in the nuclear force.
  By numerically evaluating the modified OPEP incorporating its spin and isospin
  dependencies,
  we show that the range of
  the potential decreases in both directions parallel and perpendicular to the magnetic
  field as the field strength increases.
  We also compute the resulting energy shift of the deuteron due to the modified OPEP,
  which can reach the order of 1\,MeV around $|eB| = m_\pi^2$, which is comparable to the deuteron binding energy.
\end{abstract}

\subjectindex{B60 D00 D01 D03}

\maketitle

\section{Introduction}

The Yukawa theory of the pion and the nuclear force laid the foundation for modern particle physics~\cite{Yukawa:1935xg}.
Even today, pions---understood as pseudo-Nambu-Goldstone bosons arising from the spontaneous breaking of chiral symmetry~\cite{Nambu:1961tp}---remain central to the modern theory of nuclear force~\cite{Weinberg:1990rz,Weinberg:1991um,Ordonez:1993tn,Ordonez:1995rz} and lead to the development of the chiral effective field theory ($\chi$EFT) for nucleons and pions (see, e.g., Refs.~\cite{Bedaque:2002mn,Epelbaum:2008ga,Machleidt:2011zz,Hammer:2019poc} for reviews).
Thanks to $\chi$EFT, nuclear potentials at low energies can now be computed systematically, whereas they were previously derived phenomenologically by fitting experimental data.
Moreover, recent theoretical advances have enabled first-principles derivations of nuclear force from lattice quantum chromodynamics (lattice QCD)~\cite{Ishii:2006ec,Aoki:2009ji,Ishii:2012ssm}, which have significantly deepened our understanding of hadronic properties, including the possibility of dibaryon states~\cite{Aoki2020}.

Meanwhile,
strong magnetic fields are expected to be created in relativistic heavy-ion collisions at Relativistic Heavy-Ion Collider (RHIC) and the Large Hadron Collider (LHC) and also in astrophysical objects such as magnetars~\cite{Harding:2006qn,Hattori:2016emy}, motivating extensive studies of hadrons and QCD matter under intense magnetic fields (see Refs.~\cite{Miransky:2015ava,Hattori:2023egw,Adhikari:2024bfa} for review).
Notable examples include the (inverse) magnetic catalysis and its impact on chiral symmetry~\cite{Gusynin:1994xp,DElia:2011koc,Bali:2012zg,Fukushima:2012kc,Bruckmann:2013oba,DElia:2018xwo}, as well as modifications of hadron spectra and their internal structures under strong fields~\cite{Tiburzi:2008ma,Alford:2013jva,Taya:2014nha,Hattori:2015aki,Bali:2017ian,Hattori:2019ijy}.

Most previous studies have focused on single-particle properties of hadrons (such as mass spectra) or on mean-field analyses of QCD matter in background magnetic fields.
Going beyond such effective one-body treatments, it is natural to ask how hadron–hadron interactions are modified in strong magnetic fields.

Recent theoretical efforts have begun to explore how external magnetic fields modify the nuclear force.
In particular, weak-field $\mathcal{O}((eB)^2)$ corrections to the Yukawa potential {for the one pion exchange between two constituent quarks} have been evaluated~\cite{Braghin:2023ykz}.
It shows predominantly isotropic effects with small anisotropies, though the hadronic-level OPEP with its full spin–isospin operator structure remains unexplored.
Within $\chi$EFT, the OPEP has been formulated in an operator form~\cite{Deshmukh:2018owh}; however, it neither projects onto definite spin-isospin channels nor accounts for magnetic-field-induced modifications of observables such as the deuteron binding energy.

In this paper, we derive OPEP in the presence of a strong, homogeneous magnetic field.
Our analysis is based on the leading-order $\chi$EFT, or equivalently, chiral perturbation theory with heavy (nonrelativistic) nucleons.
We introduce a notion of a gauge-invariant OPEP, expressed in terms of pion Green’s functions in a magnetic field.
We then project the resulting potential onto definite spin and isospin channels, such as $T=0, S=1$ and $T=1, S=0$,
to analyze its channel-dependent behavior under the magnetic field.
In addition, we discuss its implications for the deuteron, the only two-nucleon bound state in vacuum.
We find that the range of the nuclear force mediated by charged-pion exchange decreases in both directions parallel and perpendicular to the magnetic field.
We also estimate the energy shift of deuteron using the first-order perturbation theory.

The organization of this paper is as follows.
In Sec.~\ref{sec:Lagrangian}, we review chiral perturbation theory with nonrelativistic nucleons in the presence of a background magnetic field.
Section~\ref{sec:potential} presents the derivation of gauge-invariant OPEP under a strong homogeneous magnetic field.
After demonstrating the OPEP in various channels in Sec.~\ref{sec:OPEP}, we then discuss its implications to deuteron in Sec.~\ref{sec:application}.
Finally, we summarize our results and discuss potential outlooks in Sec.~\ref{sec:Summary}.
Appendices~\ref{sec:Green's function} and \ref{sec:potential-matrix-element} provide the derivations of charged-pion Green’s functions and the spin-dependent matrix elements for the obtained OPEP, respectively.

\section{Leading-order $\chi$EFT under magnetic field}
\label{sec:Lagrangian}

Our analysis relies on $\chi$EFT; namely, chiral perturbation theory that describes the interaction of pions with nucleons~\cite{Weinberg:1968de,Coleman:1969sm,Callan:1969sn}.
Based on the nonlinear realization of chiral symmetry, the leading-order chiral Lagrangian in the presence of a background electromagnetic field is given by~(see, e.g., Ref.~\cite{Scherer:2012xha})
\begin{align}
  \Lcal
  = & \frac{f_\pi^2}{4} \eta^{\mu\nu}
  \tr \left( D_\mu U D_\nu U^{\dag} \right)
  + \frac{f_\pi^2 \Bcal}{2}
  \tr \left( M^\dag U + U^\dag M \right)
  \nonumber                           \\
    & + \pNcal
  \left[
    \rmi \gamma^\mu D_\mu
    - m_N
    + g_A \gamma^\mu \gamma_5 \tilalpha_{\mu\perp}
    \right] \Ncal ,
  \label{eq:Lagrangian-pi-N-1}
\end{align}
where $\gamma^\mu$'s are the gamma matrices.
We parametrize the matrix-valued field $U := \rme^{\rmi \pi^a \tau_a/f_\pi}$ with the pion fields $\pi^a$ and the Pauli matrices $\tau_a~ (a=1,2,3)$ acting on the isospin space.
Throughout the paper, we adopt the mostly-minus convention for the Minkowski metric $\eta_{\mu\nu} = \diag (+1,-1,-1,-1)$.

Assuming the approximate $\SU(2)_R \times \SU(2)_L$ chiral symmetry, the quark-mass matrix takes the form $M = \diag (m,m)$, where $m = m_u = m_d$ is the current quark mass for the up and down quarks.
The leading-order pion Lagrangian involves two low-energy coefficients, $f_\pi$ and $\Bcal$, which are related to the pion-decay constant and the chiral condensate, respectively.
We also introduce covariant derivatives as
\begin{subequations}
  \begin{align}
    D_\mu U
     & :=
    \partial_\mu U + \frac{\rmi}{2} e A_\mu (\tau_3 U - U \tau_3),
    \\
    D_\mu \Ncal
     & :=
    \partial_\mu \Ncal
    + \rmi \tilalpha_{\mu\para} \Ncal
    + \frac{\rmi}{2} e A_\mu \Ncal ,
  \end{align}
\end{subequations}
where $A_\mu$ is the background $\U(1)$ (electromagnetic) gauge field.
The second line of Eq.~\eqref{eq:Lagrangian-pi-N-1} describes the contribution from the relativistic nucleon field $\Ncal = (p_D,n_D)^\top$, consisting of the proton and neutron fields $p_D$ and $n_D$, both of which are four-component Dirac spinors, with the nucleon mass $m_N$ and axial-vector coupling $g_A$.
These nucleons interact with pions via the gauged Maurer-Cartan one-form,
\begin{align}
  \begin{split}
    \tilalpha_{\mu \para}
     & := \frac{1}{2 \rmi}
    \left[
      \xi \partial_\mu
      \xi^{-1}
      + \xi^{-1} \partial_\mu
      \xi
      + \frac{\rmi}{2} e A_\mu
      ( \xi \tau_3 \xi^{-1} + \xi^{-1} \tau_3 \xi)
      \right],
    \\
    \tilalpha_{\mu \perp}
     & := \frac{1}{2 \rmi}
    \left[
      \xi \partial_\mu
      \xi^{-1}
      - \xi^{-1} \partial_\mu
      \xi
      + \frac{\rmi}{2} e A_\mu
      ( \xi \tau_3 \xi^{-1} - \xi^{-1} \tau_3 \xi)
      \right],
  \end{split}
\end{align}
with the coset element $\xi := \rme^{\rmi \pi^a \tau_a/2f_\pi} \in \SU(2)_R \times \SU(2)_L/\SU(2)_V$.

To compute OPEP, we simplify the chiral Lagrangian~\eqref{eq:Lagrangian-pi-N-1} by expanding $U =\rme^{\rmi \pi^a \tau_a/f_\pi}$ with respect to the pion fields $\pi^a$ and by taking the heavy-baryon (nonrelativistic) limit for the nucleon field.
As a result, we obtain the following leading-order effective Lagrangian for pions and nonrelativistic (two-component) nucleon field $N = (p,n)^\top$:
\begin{align}
  \Lcal_{\eff}
   & = \eta^{\mu\nu}D^+_\mu \pi^+ D^-_\nu \pi^-
  - m_\pi^2 \pi^+ \pi^-
  + \frac{1}{2} (\partial_\mu \pi^0)^2
  - \frac{1}{2} m_\pi^2 (\pi^0)^2 \nonumber     \\
   & \hspace{120pt} + N^\dag \rmi D_0 N
  - \frac{g_A}{2f_\pi} \sum_{a=0,\pm}D_i^a \pi^a N^\dag \sigma^i \tau_a N
  ,
  \label{eq:effective-Lagrangian}
\end{align}
with the pion mass squared $m_\pi^2 := 2 m \Bcal$, the spin Pauli matrices $\sigma^i~(i=1,2,3)$, and the covariant derivative for nucleons,
\begin{align}
  D_\mu N := \partial_\mu N + \rmi e A_\mu \frac{ \tau_0 + I_{2\times 2}}{2} N.
\end{align}
Note that the covariant derivative contains the gauge field $A_\mu$, through which the $NN\pi$ vertex acquires the dependence on the external gauge field.
We use an isospin basis defined by\footnote{%
  With our normalization in Eq.~\eqref{eq:isospin-matrices}, the product of isospin operators is decomposed as $\bm{\tau}^{(1)}\!\cdot\!\bm{\tau}^{(2)} = \tau^{(1)}_+ \tau^{(2)}_- + \tau^{(1)}_- \tau^{(2)}_+ + \tau^{(1)}_0\tau^{(2)}_0$.  It differs from another common convention $\tau'_\pm :=(\tau_1\pm\rmi\tau_2)/2$, for which $\bm{\tau}^{(1)}\!\cdot\!\bm{\tau}^{(2)} = 2\tau^{(1)\prime}_{+}\tau^{(2)\prime}_{-} + 2\tau^{(1)\prime}_{-}\tau^{(2)\prime}_{+} + \tau^{(1)}_0\tau^{(2)}_0$.
}
\begin{align}
  \tau_+ := \frac{\tau_1 + \rmi \tau_2}{\sqrt{2}} =
  \begin{pmatrix}
    0 & \sqrt{2} \\
    0 & 0
  \end{pmatrix}
  , \quad
  \tau_- := \frac{\tau_1 - \rmi \tau_2}{\sqrt{2}} =
  \begin{pmatrix}
    0        & 0 \\
    \sqrt{2} & 0
  \end{pmatrix}
  , \quad
  \tau_0 := \tau_3 =
  \begin{pmatrix}
    1 & 0  \\
    0 & -1
  \end{pmatrix} ,
  \label{eq:isospin-matrices}
\end{align}
for which the covariant derivatives for the pion fields are%
\footnote{
  We adopt a slightly redundant notation by including the $\pm$ and $0$ indices on the covariant derivatives, as this will be convenient in the subsequent discussion.
}
\begin{align}
  D^\pm_\mu \pi^{\pm}
  := (\partial_\mu  \pm \rmi e A_\mu) \pi^{\pm},
  \quad
  D^0_\mu \pi^0 := \partial_\mu \pi^0. \label{eq:covariant-derivatives-pion}
\end{align}
We note that the leading two-pion--nucleon coupling (the Weinberg-Tomozawa term) is omitted here, as it does not contribute to OPEP that we aim to compute.

As a final remark, we clarify the power counting scheme employed in this paper.
We count the momentum $p_\mu$ of pions and nonrelativistic nucleons and the background gauge field $eA_\mu$ as $p_\mu=\mathcal{O}(\epsilon)$ and $eA_\mu=\mathcal{O}(\epsilon)$, respectively, while the quark-mass matrix is counted as $M = \mathcal{O}(\epsilon^2)$.
This counting allows us to take magnetic-field strengths as large as the pion-mass scale, $|e\bm{B}| \sim m_\pi^2 = \mathcal{O}(\epsilon^2)$.
In this scheme, higher-order terms of $\mathcal{O}(\epsilon^{4})$ or above lie beyond the accuracy of our leading-order analysis and are neglected.
As a rough estimate of the ultraviolet cutoff $\Lambda_{\mathrm{UV}}$, we take $\Lambda_{\mathrm{UV}} \simeq m_\rho = 770\,$MeV.
Accordingly, we numerically consider magnetic-field strengths up to
$|e\bm{B}| \lesssim m_\rho^2 \sim 30 m_\pi^2$.%
\footnote{\label{footnote2}
  This estimate is admittedly crude.
  For instance, it neglects the mass shifts of other hadrons due to the applied magnetic field~\cite{Adhikari:2024bfa}.
  A more realistic estimate would take into account the mass shift of, for example, the $\rho$ meson, which we do not consider here.
}
For magnetic fields approaching the upper end of this range,
higher-order contributions beyond $\mathcal{O}(\epsilon^2)$ may become quantitatively important, and the present leading-order analysis should be regarded as subject to truncation uncertainties of the chiral expansion.

In summary, Eq.~\eqref{eq:effective-Lagrangian} captures the leading-order contribution to the nuclear force from the OPEP in the presence of a magnetic field.
Within our power-counting scheme, only the charged pions couple minimally to the electromagnetic field.
We thus concentrate on the charged-pion exchange under a homogeneous magnetic background in what follows.

\section{One-pion exchange potential in
  magnetic field}
\label{sec:potential}

In this section, we derive the one-pion exchange potential (OPEP) in the presence of a magnetic field.
In Sec.~\ref{sec:solve-eom}, we solve the equations of motion for the pion fields and obtain the corresponding effective interaction Hamiltonian between two nucleons.
In Sec.~\ref{sec:gauge-inv}, we introduce a gauge-invariant definition of OPEP.
Finally, in Sec.~\ref{sec:Limitting-behavior}, we present simplified analytic expressions of the OPEP in weak- and strong-field limits.

\subsection{Deriving two-nucleon interaction Hamiltonian}
\label{sec:solve-eom}

We derive OPEP by solving the equation of motion for the pion fields in a static setup where the source term is supplied by a test nucleon at rest.
By substituting the resulting static pion solution into the interaction Hamiltonian,
\begin{align}
  H_{\mathrm{int}} = \frac{g_A}{2f_\pi}
  \sum_{a= 0,\pm} \int & \diff^3{x} D_i^a \pi^a N^\dag \sigma^i \tau_a N ,
\end{align}
we obtain the effective interaction Hamiltonian between two nucleons.
The resulting effective Hamiltonian allows us to determine the induced potential between the two nucleons.

Let us solve the equation of motion for a charged-pion field in a homogeneous static magnetic field directed along the $z$-axis: $e\bB = eB\hat{\bz} = (0,0,eB)$.
From the leading-order effective Lagrangian~\eqref{eq:effective-Lagrangian}, we obtain the classical equation of motion for the pion field as
\begin{align}
  \left[\eta^{\mu\nu} D_\mu^a D_\nu^a + m_\pi^2 \right] \pi^a
  = \frac{g_A}{2f_\pi} D_i^{a} \bigl[N^\dag \sigma^i \tau_{-a} N\bigr],
\end{align}
where the unsummed indices $a=0,\pm$ denote the electric charge of the pion field, and $\tau_{-a}$ is the corresponding isospin Pauli matrix, as defined in Eq.~\eqref{eq:isospin-matrices}.
In the following analysis, we focus on the static case and neglect the temporal derivatives of the pion field.
Introducing a coordinate-space Green's function for pions in the presence of an external gauge field, $G^{a}({\bx,\bx'}|A)$, such that
\begin{align}
  \bigl[-\delta^{ij} D_i^a D_j^a + m_\pi^2 \bigr]
  G^{a}({\bx,\bx'}|A)=\delta^{(3)} ({\bx-\bx'}),
  \label{eq:propagator-spacerep}
\end{align}
we find that the induced pion field, sourced by the test nucleon, is given by
\begin{align}
  \pi^a ({\bx}) =
  -\frac{g_A}{2f_\pi}\int \diff^3{x'}
   & \Bigl[\bigl(D_i^{-a}\bigr)_{{x'}} G^{a}({\bx,\bx'}|A)\Bigr]
  N^\dag({\bx'})\sigma^i\tau_{-a} N({\bx'}).
  \label{eq:induced-pion}
\end{align}
Here the subscript ${{x'}}$ of $\bigl(D_i^{-a}\bigr)_{{x'}}$ indicates that the covariant derivative acts on the coordinate ${\bx'}$.
We can express the pion Green's function $G^{a}({\bx,\bx'}|A)$ in a more explicit form.
For the neutral pion ($a=0$), it does not couple to an electromagnetic field, so its Green's function retains the familiar Yukawa form:
\begin{align}
  G^0( {\bx,\bx'}|A) = \frac{1}{4\pi} \frac{\rme^{- m_{\pi} |{\bx - \bx'}|}}{|{\bx - \bx'}|}.
  \label{eq:neutral-pion-propag-space}
\end{align}
In contrast, the charged-pion Green's function ($a = \pm$) is significantly modified by the magnetic field when $e B/m_\pi^2 = \mathcal{O}(1)$.
To incorporate the magnetic-field effects without relying on weak-field expansion, we employ Schwinger's proper-time method~\cite{Schwinger:1951nm}, which allows for an all-order treatment of the applied background field (see, e.g., Ref.~\cite{Hattori:2023egw} for a recent review).
The explicit expression for $G^{a}({\bx,\bx'}|A)$ in an arbitrary gauge is given by (see Appendix~\ref{sec:Green's function} for details):
\begin{align}
  \begin{split}
    G^\pm({\bm{x},\bm{x}'}|A)
     & = e^{\pm \rmi\,{\Phi_A({\bm{x},\bm{x}')}}}\;
    G_B({\bm{x}-\bm{x'}}),
    \\
    G_B({\bm{x}-\bm{x}'})
     & =\frac{|eB|}{8\pi^2}\int_0^\infty \diff s\frac{1}{\sinh(|eB|s)}\sqrt{\frac{\pi}{s}}
    \exp \left(-m_\pi^2s-\frac{1}{4s}(z-z')^2-\frac{|eB|}{4}\frac{|{\bx_\perp-{\bx'_\perp}}|^2}{\tanh(|eB|s)}\right),
    \\
    \Phi_A({\bx,\bx'})
     & = -\,e{\int_{\bm{x}'}^{\bm{x}}}
    \diff\bm{{\xi}}\cdot
    \Bigl[\bm{A}(\bm{{\xi}})+\tfrac12(\bm{{\xi}}-{\bm{x}'})\times\bm{B}\Bigr]\,,
  \end{split}
  \label{eq:charge-pion-propag-space}
\end{align}
where $z$ and $z'$ denote the coordinates along the direction of the magnetic field, and {$\bx_\perp$ and $\bx_\perp'$} denote the transverse components.
We also introduced the Green’s function $G_B({\bm{x}-\bm{x}'})$,
which is evaluated in the Fock–Schwinger symmetric (FS) gauge.
The corresponding gauge potential $A_i^{\text{FS}}$ in the FS gauge is defined by $({x^i-x'^{\,i}})A_i^{\text{FS}}=0$,
leading to $A_i^{\text{FS}}=-\tfrac{1}{2}F_{ij}({x^j-x'^{\,j}})$.
For a homogeneous magnetic field $\bm{B}=B\hat{\bm{z}}$, this reduces to
\begin{align}
  \bm{A}^{\text{FS}}({\bx})
  = -\frac{1}{2}\, ({\bx - \bx'})\times \bm{B}
  = \left(-\frac{B}{2}(y-y'),\, \frac{B}{2}(x-x'),\, 0\right).
  \label{eq:Fock-Schwinger-gauge}
\end{align}
The phase factor $\Phi_A({\bx,\bx'})$ is the Schwinger phase, which governs the gauge-transformation properties of the Green’s function:
\begin{align}
  \begin{cases}
    G^+ ({\bm{x},\bm{x}'}|A)
    \to \rme^{\rmi e {(\alpha(\bx)-\alpha (\bx'))}} G^+ ({\bm{x},\bm{x}'}|A'),
    \\
    G^- ({\bm{x},\bm{x}'}|A)
    \to \rme^{-\rmi e { ( \alpha(\bx) - \alpha (\bx'))}} G^- ({\bm{x},\bm{x}'}|A'),
  \end{cases}
  \label{eq:gauge-transformation-G}
\end{align}
where the gauge field transforms as $A_\mu \to A_\mu' = A_\mu - \partial_\mu \alpha$.

Finally, we introduce the gauge-invariant two-nucleon interaction, $H_\text{int}$, describing the one-pion-exchange-mediated force. It follows from substituting the induced pion field  Eq.~\eqref{eq:induced-pion} into the interaction term in Eq.~\eqref{eq:effective-Lagrangian}, yielding
\begin{align}
  H_\text{int}
  =-\frac{g_A^2}{4f_\pi^2}\sum_{a=0,\pm}
  \int & \diff^3{x} \diff^3{x'}
  N^\dagger({\bx'})\sigma^j \tau_{-a}N({\bx'})
  \nonumber                     \\
       & \hspace{40pt}\times
  \Bigl[(D_i^{a})_{x}(D_j^{-a})_{{x'}} G^{a}({\bx,\bx'}|A)\Bigr]
  N^\dagger({\bx})\sigma^i\tau_a N({\bx})\,.
  \label{eq:interacting-Hamiltonian}
\end{align}

\subsection{Definition of gauge-invariant potential}
\label{sec:gauge-inv}

The Hamiltonian~\eqref{eq:interacting-Hamiltonian} encodes all the information about the interaction between two nucleons induced by the one-pion exchange.
Physical observables calculated from this Hamiltonian, such as scattering cross sections and binding energies, are gauge independent.
On the other hand, the OPEP itself is not an observable; it is an intermediate, gauge-dependent quantity, whose definition requires a specific prescription.

In principle, any gauge choice can be used to define such a potential as long as observables remain unaffected.
In this work, we adopt a manifestly gauge-invariant definition by introducing the Wilson line~\cite{Deshmukh:2018owh}.
The Wilson line cancels the gauge-dependent Schwinger phase associated with the charged-pion propagator, ensuring that the same potential is obtained in any gauge.
In particular, in the Fock–Schwinger (FS) gauge the Wilson line becomes unity, so the potential can be evaluated without explicit reference to the Schwinger phase.
This prescription allows us to visualize the OPEP as a unique function of the relative coordinate, which should facilitate comparisons with other approaches.

Let us now detail the formulation and demonstrate that this gauge-invariant construction reduces to the expression in the FS gauge.
We choose a reference position $\bx_0$ and introduce a ``dressed" nucleon field $\tilN(\bx)$ as
\begin{align}
  \tilN (\bx) := W(\bx,\bx_0) N(\bx),
  \label{eq:wilsoned-nucleon_field}
\end{align}
where the spatial Wilson line $W(\br,\br_0)$ is defined as
\begin{align}
  W(\bx,\bx_0) := \exp\left(\rmi e \frac{\tau_0+I_{2\times2}}{2} \int_{\bx_0}^{\bx} \bm{A} (\bm{\xi})\cdot\diff \bm{\xi} \right) .
\end{align}
Due to the dressing, a nucleon state is now produced by $\tilde{N}$, not by $N$, as
\begin{align}
  \ket{\tilde{N}(\bx)} := \tilN^\dagger(\bx) \ket{0} , \label{eq:dessed_state}
\end{align}
and multi-nucleon states are produced similarly by acting $\tilde{N}$ multiple times.
Since the Wilson line transforms under a gauge transformation as
\begin{align}
  W(\bx,\bx_0) \to
  \rme^{- \rmi e \frac{\tau_0+I_{2\times2}}{2} [ \alpha (\bx) - \alpha (\bx_0)]} W(\bx,\bx_0) ,
\end{align}
a nucleon state $\ket{\tilde{N}}$ transforms as
\begin{align}
  \ket{\tilde{N}(\bx)} \to
  \rme^{-\rmi e \frac{\tau_0+I_{2\times2}}{2} \alpha (\bx_0)} \ket{\tilde{N}(\bx)} ,
\end{align}
i.e., a proton or nucleon state transforms with a single phase factor at the reference point $\bx_0$ as $\ket{\tilde{p}(\bx)} \to \rme^{-\rmi e \alpha (\bx_0)} \ket{\tilde{p}(\bx)}$ and $\ket{\tilde{n}(\bx)} \to \ket{\tilde{n}(\bx)}$.  Accordingly, for any set of initial and final states with the dressed nucleons, which we write $\ket{\tilde{i}}$ and $\bra{\tilde{f}}$, respectively, the phase factors that they acquire after a gauge transformation are, respectively, $\rme^{-\rmi e n \alpha (\bx_0)}$ and $\rme^{+\rmi e n' \alpha (\bx_0)}$, where $n$ and $n'$ are the numbers of protons contained in these states.
Since the total electric charge is conserved, we have $n=n'$ and understand that the phase factors coming from $\ket{\tilde{i}}$ and $\bra{\tilde{f}}$ cancel with each other, making the matrix element $\bra{\tilde{f}} H_\text{int} \ket{\tilde{i}}$ gauge invariant.

We are in a position to derive a gauge-invariant OPEP from the Hamiltonian~\eqref{eq:interacting-Hamiltonian}.
To do so, we first rewrite $H_\text{int}$~\eqref{eq:interacting-Hamiltonian} with the dressed nucleon field, which yields
\begin{align}
  \begin{split}
    H_\text{int}
     & =-\frac{g_A^2}{4f_\pi^2} \sum_{a=0,\pm}
    \int \diff^3{x}\,\diff^3{x}'
    \tilN^\dagger({\bx'})\sigma^j \tau_{-a}\tilN({\bx'}) \\
     & \hspace{60pt}\times
    \rme^{+\rmi a e\int_{{\bx'}}^{{\bx}} \bm{A}(\bm{\xi})\cdot\diff \bm{\xi}}
    \Bigl[ (D_i^{a})_{x}(D_j^{-a})_{{x'}} G^{a} ({\bx,\bx'}|A) \Bigr]
    \tilN^\dagger({\bx})\sigma^i\tau_a \tilN({\bx}) ,
  \end{split}
  \label{eq:Hint_with_tilde_N_and_phase}
\end{align}
where the Wilson-line factor $ \rme^{+\rmi a e\int_{\bx'}^{\bx} \bm{A}(\bm{\xi})\cdot\diff \bm{\xi}}$ arises due to the multiplication of $W$ and $W^\dagger$'s from the dressed nucleon fields.
It is worth noting that the additional Wilson-line factor makes the kernel in the Hamiltonian to be the one simply evaluated in the Fock-Schwinger gauge:
\begin{align}
  \rme^{+\rmi a e\int_{\bx'}^{\bx} \bm{A}(\bm{\xi})\cdot\diff \bm{\xi}}
  \Bigl[ (D_i^{a})_x(D_j^{-a})_{x'} G^{a} (\bx,\bx'|A)\Bigl]=
  (D_i^{a})_x^\text{FS}(D_j^{-a})_{x'}^\text{FS}
  G_B(\bm{x}-\bm{x'}) ,
\end{align}
where the covariant derivatives with a superscript $``\text{FS}"$ are
\begin{align}
  \begin{split}
    (D_i^{a})^{{\text{FS}}}= \partial_i - \text{i} a e \frac{1}{2}\,[\,(\bm{x}-\bm{x}')\times\bm{B}\,]_i.
  \end{split}
\end{align}

We then define OPEP for all possible one-pion-exchange channels via the corresponding matrix elements as
\begin{align}
  \begin{split}
    V_{0}(\br_1-\br_2)\braket{\tilde{p}(\br_3)\tilde{p}(\br_4)|\tilde{p}(\br_1)\tilde{p}(\br_2)}  & := \frac{1}{2}\bra{\tilde{p}(\br_3)\tilde{p}(\br_4)} H_\text{int} \ket{\tilde{p}(\br_1)\tilde{p}(\br_2)}            \\
                                                                                                  & \hspace{3pt}=\frac{1}{2}\bra{\tilde{n}(\br_3)\tilde{n}(\br_4)} H_\text{int} \ket{\tilde{n}(\br_1)\tilde{n}(\br_2)},
    \\
    V_{pn}(\br_1-\br_2)\braket{\tilde{n}(\br_3)\tilde{p}(\br_4)|\tilde{n}(\br_1)\tilde{p}(\br_2)} & := \frac{1}{4}\bra{\tilde{n}(\br_3)\tilde{p}(\br_4)} H_\text{int} \ket{\tilde{p}(\br_1)\tilde{n}(\br_2)}     ,      \\
    V_{np}(\br_1-\br_2)\braket{\tilde{p}(\br_3)\tilde{n}(\br_4)|\tilde{p}(\br_1)\tilde{n}(\br_2)} & := \frac{1}{4}\bra{\tilde{p}(\br_3)\tilde{n}(\br_4)} H_\text{int} \ket{\tilde{n}(\br_1)\tilde{p}(\br_2)}.
  \end{split}
\end{align}
The factor \(1/2\) in \(V_0\) avoids double counting of equivalent contributions that arise when defining the potential as a matrix element of \(H_\text{int}\).
Similarly, the factors \(1/4\) in \(V_{pn}\) and \(V_{np}\) consist of two \(1/2\) factors:
one accounts for double counting of equivalent charged-pion exchange processes,
and the second factor $1/2$ compensates for the additional factor of 2 that arises when
$N^\dagger \tau_\pm N$ is rewritten in terms of the proton and neutron fields
(e.g., $N^\dagger \tau_- N = \sqrt{2}\,n^\dagger p$), ensuring consistency with $V_0$.
It is easy to show via direct calculations
\begin{align}
  \begin{split}
    V_{0} (\br_1-\br_2)
     & := -\frac{g_A^2}{4f_\pi^2}
    \sigma^i_{(1)}\sigma^j_{(2)}
    (\partial_i)_{x} (\partial_j)_{{x'}}
    G^0({\bx-\bx'})\big|_{{\bx=\br_1,\bx'=\br_2}} ,
    \\
    V_{pn}(\br_1-\br_2)
     & := -\frac{g_A^2}{8f_\pi^2}
    \Bigl[
    \sigma^i_{(1)}\sigma^j_{(2)}
    \bigl(D_i^{-}\bigr)^\text{FS}_{x} \bigl(D_j^{+}\bigr)^\text{FS}_{{x'}}
    G_B({\bx-\bx'})\big|_{{\bx=\br_1,\bx'=\br_2}}
    \\
     & \hspace{80pt}
    +\sigma^j_{(1)}\sigma^i_{(2)}
    \bigl(D_i^{+}\bigr)^\text{FS}_{x} \bigl(D_j^{-}\bigr)^\text{FS}_{{x'}}
    G_B({\bx-\bx'})\big|_{{\bx=\br_2,\bx'=\br_1}}
    \Bigr]
    ,
    \\
    V_{np}(\br_1-\br_2)
     & :=V^\dagger_{pn}(\br_1-\br_2).
  \end{split}
  \label{eq:potential-component}
\end{align}

Physically, $V_{pn}$ ($V_{np}$) accounts for the contribution to the $pn$ ($np$) channel, i.e., a charged-pion-exchange process in which a proton at a certain position emits one $\pi^+$ (absorbs one $\pi^-$) and a neutron absorbs (or emits) it.
The other one, $V_0$, is responsible for the $pp$ and $nn$ channels, which are mediated by a neutral pion.
The obtained OPEP is clearly modified in the $pn$ and $np$ channels, while it remains unchanged in the $pp$ and $nn$ channels.
This is the result of the breaking of the isospin and spin symmetries by a magnetic field.

For later calculations, it is convenient to introduce $V^{B\neq0}_\text{OPEP}$ an effective OPEP that incorporates all the information from each channel’s OPEP in Eq.~\eqref{eq:potential-component} as
\begin{align}
  \begin{split}
    V^{B\neq0}_\text{{OPEP}}(\br)
     & :=
    V_{pn}({\br})\,\tau^{(1)}_-\tau^{(2)}_+
    + V_{np}({\br})\,\tau_+^{(1)}\tau_-^{(2)}
    + V_{0} ({\br}) \,\tau^{(1)}_0 \tau^{(2)}_0,
  \end{split}
  \label{eq:one_charged_pion_exchange_potential}
\end{align}
where we introduced the relative coordinate $\br = \br_1 - \br_2$.
The operators $\sigma^{i}_{(1)}$ and $\sigma^{i}_{(2)}$ [$\tau_{\pm}^{(1)}$ and $\tau_{\pm}^{(2)}$] act on the spin (isospin) spaces of the nucleons located at $\br_1$ and $\br_2$, respectively.

\subsection{Weak and strong magnetic-field limits}
\label{sec:Limitting-behavior}

Here, we consider two limiting cases of the OPEP in a magnetic field, where simple analytic expressions are available: the weak-field limit ($|eB| \ll m_\pi^2$) and the strong-field limit ($|eB| \gg m_\pi^2$).%
\footnote{Our OPEP in a magnetic field~\eqref{eq:one_charged_pion_exchange_potential} is applicable to the strong-field region such that $eB \gtrsim m_\pi^2$, so long as the validity condition $eB \lesssim \Lambda_{\rm UV}^2 \sim 30m_\pi^2$ is satisfied [see discussion below Eq.~\eqref{eq:covariant-derivatives-pion}]. }
\medskip
\paragraph{Weak magnetic-field limit}
When the applied magnetic field is sufficiently weak compared to the pion-mass scale in vacuum, we can perform a perturbative expansion with respect to the small parameter $|eB|/m_\pi^2 \ll 1$.
Expanding the full charged-pion propagator~\eqref{eq:charge-pion-propag-space}, we obtain
\begin{align}
  \begin{split}
    G_B({\bx-\bx'})
     & =\frac{1}{4\pi}\frac{1}{|{\bx-\bx'}|}\rme^{-m_\pi|{\bx-\bx'}|}-\frac{|eB|^2}{96\pi m_\pi^3}(1+m_\pi|{\bx-\bx'}|+m_\pi^2|{\bx_\perp-\bx_\perp'}|^2)\rme^{-m_\pi|{\bx-\bx'}|} \\
     & \hspace{300pt}+ \mathcal{O} \left( |eB|^4 \right).
  \end{split}
\end{align}
Using this weak-field expression up to $\mathcal{O}(|eB|^2)$, we find that the weak magnetic-field dependence of the OPEPs, $V_{pn}$ and $V_{np}$, is given by
\begin{align}
  \begin{split}
    {\lim_{eB\ll m_\pi^2}}V_{pn} (\br)
     & = {\lim_{eB\ll m_\pi^2}}\left[V_{np}(\br)\right]^\dagger                                                                                                                                      \\
     & =\frac{g_A^2}{16\pi f_\pi^2} \left[
    \left(\frac{1}{r^2}+\frac{m_\pi}{r}+\frac{m_\pi^2}{3}\right)S_{12}
    +\frac{m_\pi^2}{3}\bm{\sigma}_{(1)}\cdot\bm{\sigma}_{(2)}
    \right] \frac{\mathrm{e}^{-m_\pi r}}{r}                                                                                                                                                          \\
     & \hspace{10pt} + \frac{g_A^2 |eB|^2}{384\pi m_\pi^2 f_\pi^2} \Bigg[\left(1+m_\pi r-m_\pi^2\br_\perp^2\right)\bm{\sigma}_{(1)}\cdot\bm{\sigma}_{(2)}                                            \\
     & \hspace{70pt}-\left(1+m_\pi r-m_\pi^2\br_\perp^2+3m_\pi^2r^2-m_\pi^3 r\br_\perp^2\right)\frac{(\bm{\sigma}_{(1)}\cdot \br)(\bm{\sigma}_{(2)}\cdot \br)}{r^2}                                  \\
     & \hspace{80pt}+m_\pi^2\left(\sigma^3_{(1)}z(\bm{\sigma}_{(2)}\cdot \br)+(\bm{\sigma}_{(1)}\cdot \br)\sigma^3_{(2)}z\right)+2m_\pi r\sigma^3_{(1)}\sigma^3_{(2)}\Bigg]\frac{\rme^{-m_\pi r}}{r} \\
     & \hspace{10pt}-\frac{g_A^2|eB|^2}{64\pi f_\pi^2}\left(\bm{\sigma}_{(1)}\times\br_\perp\right)_z\left(\bm{\sigma}_{(2)}\times\br_\perp\right)_z
    \frac{\rme^{-m_\pi r}}{r},
  \end{split}
  \label{eq:OPEP-pn-weak}
\end{align}
where $\br =: (\br_\perp, z)$ and $r = |\br|$.
We note that the first line of Eq.~(\ref{eq:OPEP-pn-weak}) exactly coincides with $V_0$, after evaluating the derivatives in Eq.~(\ref{eq:potential-component}), with the tensor operator,
\begin{align}
  S_{12} :=\frac{3}{r^2}(\bm{\sigma}_{(1)}\cdot\bm{r})(\bm{\sigma}_{(2)}\cdot\bm{r})
  -\bm{\sigma}_{(1)}\cdot\bm{\sigma}_{(2)}.
  \label{eq:tensor-operator}
\end{align}
As usual, the tensor operator couples the nucleon spins to their relative coordinates and results in anisotropy of OPEP for the spin-triplet ($S=1$) channel, even in the absence of magnetic fields.
The numerical prefactor for the ${\mathcal O}(|eB|^2)$ term turns out to be rather small, implying that the zero-field limit dominates even around $|eB| \sim m_\pi^2$.

\paragraph{Strong magnetic-field limit}
When the applied magnetic field is sufficiently strong, the dynamics of charged
pions is dominated by the lowest Landau level (LLL), and a similar LLL
dominance is expected to appear in the OPEP.
To clarify the conditions for this dominance, we recall that the charged-pion dynamics becomes anisotropic in a magnetic field.

The longitudinal dynamics is governed by the effective mass $\sqrt{m_\pi^2 + (2n+1)|eB|}$ of the Landau levels ($n = 0,1,2,\cdots$), whereas the transverse dynamics is controlled by the magnetic length $1/\sqrt{|eB|}$, which sets the size of the Landau orbit [see Eqs.~\eqref{eq:propagator-momentum}–\eqref{eq:final-result-FS-gauge} in Appendix~\ref{sec:Green's function}].
As a result, the LLL ($n=0$) becomes dominant along the $z$-direction when the
longitudinal separation satisfies $|z| \gg 1/\sqrt{m_\pi^2 + |eB|}$.
In contrast, the LLL dominance in the transverse direction arises when $|\br_\perp| \gg 1/\sqrt{|eB|}$.
In the strong-field regime $|eB| \gg m_\pi^2$, the longitudinal condition
also reduces to $|z| \gg 1/\sqrt{|eB|}$.

In the LLL-dominant regime, the charged-pion Green’s function can be approximated well by the contribution from the LLL [i.e., the $n=0$ mode in Eq.~\eqref{eq:Green's function using Laguerre polynomial}] as
\begin{align}
  \begin{split}
    G_B({\bx-\bx'})
    \approx
    G_B^{\mathrm{LLL}} ({\bx-\bx'})
    := \frac{|eB|}{4\pi}\rme^{-\frac{|eB|}{4}|{\bx_\perp-\bx_\perp'}|^2}\frac{\rme^{-\sqrt{m_\pi^2+|eB|}|z-z'|}}{\sqrt{m_\pi^2+|eB|}}.
  \end{split}
\end{align}
Substituting the LLL Green’s function $G_B^{\mathrm{LLL}}(\bx-\bx')$ into Eq.~\eqref{eq:potential-component}, the OPEP induced by charged-pion exchange can be written in the LLL-dominant regime as
\begin{align}
  \begin{split}
    V_{pn}^{\text{LLL}} (\br)
     & =\left[V_{np}^{\text{LLL}}(\br)\right]^\dagger                                                                                                                                                                                                \\
     & =-\frac{g_A^2|eB|}{16\pi f_\pi^2}\Bigg[\frac{|eB|}{2\sqrt{m_\pi^2+|eB|}}\bm{\sigma}^{(1)}_\perp\cdot\bm{\sigma}^{(2)}_\perp-\frac{|eB|^2}{4\sqrt{m_\pi^2+|eB|}}(\bm{\sigma}^{(1)}_\perp\cdot\br_\perp)(\bm{\sigma}^{(2)}_\perp\cdot\br_\perp) \\
     & \hspace{60pt}-\frac{|eB|}{2}\left(\sigma^3_{(1)}(\bm{\sigma}^{(2)}_\perp\cdot\br_\perp)+(\bm{\sigma}^{(1)}_\perp\cdot\br_\perp)\sigma^3_{(2)}\right)\frac{z}{|z|} -\sqrt{m_\pi^2+|eB|} \sigma^3_{(1)}\sigma^3_{(2)}
    \\
     & \hspace{70pt}
    +\frac{|eB|^2}{4\sqrt{m_\pi^2+|eB|}}\left(\bm{\sigma}_{(1)}\times\br_\perp\right)_z\left(\bm{\sigma}_{(2)}\times\br_\perp\right)_z\Bigg]
    \rme^{-\frac{|eB|}{4}\br_\perp^2-\sqrt{m_\pi^2+|eB|}|z|} .
  \end{split}
  \label{eq:OPEP-pn-LLL}
\end{align}

\section{Behavior of OPEP}
\label{sec:OPEP}

The derived formula in the previous section enables us to evaluate the matrix elements of the OPEP.
In this section, we investigate the behavior of the OPEP in each channel:
the isospin-singlet channel in Sec.~\ref{sec:OPEP-singlet}
and the isospin-triplet channel in Sec.~\ref{sec:OPEP-triplet}.

\subsection{OPEP in the isospin singlet channel}
\label{sec:OPEP-singlet}
Let us first examine the OPEP under a background magnetic field for the isospin-singlet channel.
This channel is relevant to the deuteron, a two-nucleon bound state in the isospin-singlet ($T=0$) and spin-triplet ($S=1$).
For simplicity, we denote the OPEP evaluated in the isospin-singlet channel as
$V^{B\neq0,\,T=0}_\text{{OPEP}}(\br)$ and express a spin-triplet state as $\ket{S_z = a} := \ket{S=1,S_z=a}$, with $a=\pm1 , 0$ labeling the spin magnetic quantum number.
One can explicitly evaluate various matrix elements for two deuteron states by using Eqs.~\eqref{eq:potential-component} and \eqref{eq:one_charged_pion_exchange_potential}.
For example, the matrix elements for $\ket{S_z=+1} \to \ket{S_z=+1}$ and $\ket{S_z=+0}  \to \ket{S_z=+0}$ are given, respectively, by
\begin{align}
   & \braket{S_z=+1|V^{B\neq0,\,T=0}_\text{{OPEP}}(\br)|S_z=+1}
  \nonumber                                                     \\
   & \hspace{50pt}
  = -V_0^{+1,+1}(\br)
  + \frac{g_A^2|eB|}{32\pi^2 f_\pi^2}
  \int_0^\infty \diff s
  \frac{1}{\sinh(|eB|s)}
  \sqrt{\frac{\pi}{s^3}} \left(1-\frac{1}{2s}z^2\right)\mathcal{F}(\br; s),
  \label{eq:potential-matrix-element-11}
\end{align}
and
\begin{align}
  \begin{split}
     & \braket{S_z=0|V^{B\neq0,\,T=0}_\text{{OPEP}}(\br)|S_z=0}                                                                                                                                                              \\
     & \hspace{10pt}
    = -V_0^{0,0}(\br)
    -\frac{g_A^2|eB|}{32\pi^2 f_\pi^2}\int_0^\infty \diff s\frac{1}{\sinh(|eB|s)}\sqrt{\frac{\pi}{s^3}}\left(1-\frac{1}{2s}z^2\right)\mathcal{F}(\br; s)
    \\
     & \hspace{30pt}+\frac{g_A^2|eB|^2}{32\pi^2f_\pi^2}\int_0^\infty \diff s\frac{1}{\sinh(|eB|s)}\frac{1}{\tanh(|eB|s)}\sqrt{\frac{\pi}{s}}\left(2-\frac{|eB|}{2}\frac{\br_\perp^2}{\tanh(|eB|s)}\right)\mathcal{F}(\br; s) \\
     & \hspace{40pt}+\frac{g_A^2|eB|^3}{64\pi^2f_\pi^2}\int_0^\infty \diff s\frac{\br_\perp^2}{\sinh(|eB|s)}\sqrt{\frac{\pi}{s}}\mathcal{F}(\br; s) ,
  \end{split}
  \label{eq:potential-matrix-element-00}
\end{align}
where $V^{a,b}_{0} := \bra{S_z=a} V_0 \ket{S_z=b}$, with $V_0$ being the OPEP for the neutral-pion channel~\eqref{eq:potential-component}.
For notational convenience, we define a common exponential factor as
\begin{equation}
  \mathcal{F}(\br; s) :=
  \exp\!\left(
  -m_\pi^2 s
  - \frac{z^2}{4s}
  - \frac{|eB|}{4}\frac{\br_\perp^2}{\tanh(|eB|s)}
  \right),
  \label{eq:def-F}
\end{equation}
which frequently appears in the proper-time representations of the OPEP.
In the same manner, we can explicitly write down the proper-time-integral formulas for the other $S_z$-dependent deuteron matrix elements (see Appendix~\ref{sec:potential-matrix-element} for a complete list).

\begin{figure*}[t]
  \centering
  \begin{minipage}{0.49\textwidth}
    \centering
    \adjustbox{raise=1mm}{\includegraphics[width=\textwidth]{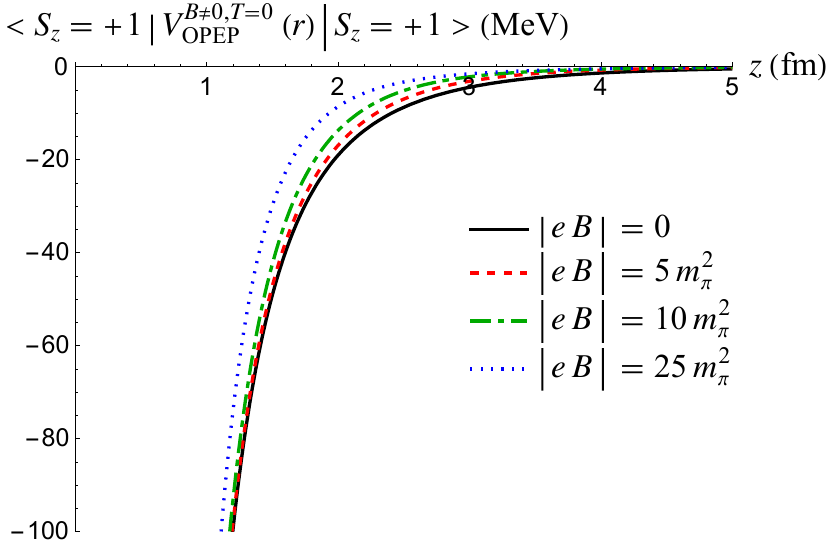}}
  \end{minipage}
  \hfill
  \begin{minipage}{0.47\textwidth}
    \centering
    \includegraphics[width=\textwidth]{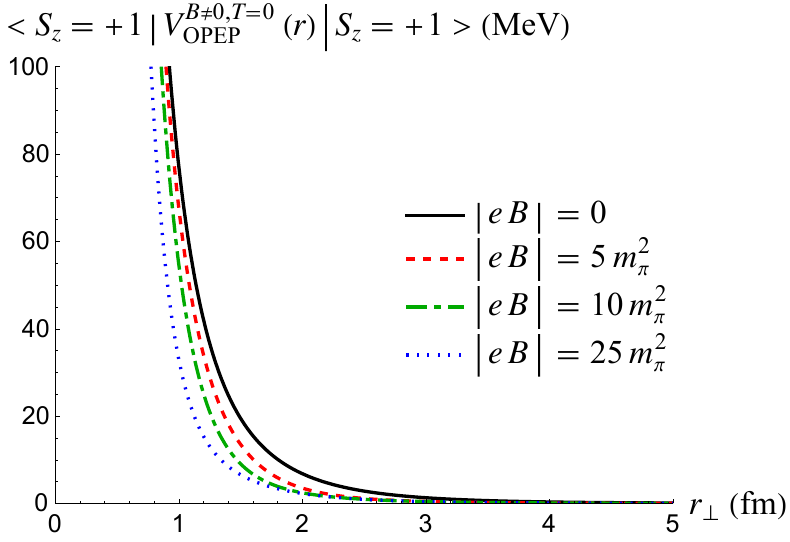}
  \end{minipage}
  \caption{
    Magnetic-field dependence of the OPEP matrix element between the $\ket{S_z=+1}$ states in the isospin-singlet ($T=0$) and spin-triplet ($S=1$) channel, as given in Eq.~\eqref{eq:potential-matrix-element-11}.
    Left panel: potential along the magnetic field (longitudinal) direction;
    right panel: potential along the perpendicular (transverse) direction.
    For the plots, we use the vacuum values: $g_A=1.27,\; f_\pi=92\text{ MeV}$, and $m_\pi=138\text{ MeV}$.
  }
  \label{fig:V11}
\end{figure*}

\begin{figure*}[t]
  \centering
  \begin{minipage}{0.49\textwidth}
    \centering
    \includegraphics[width=\textwidth]{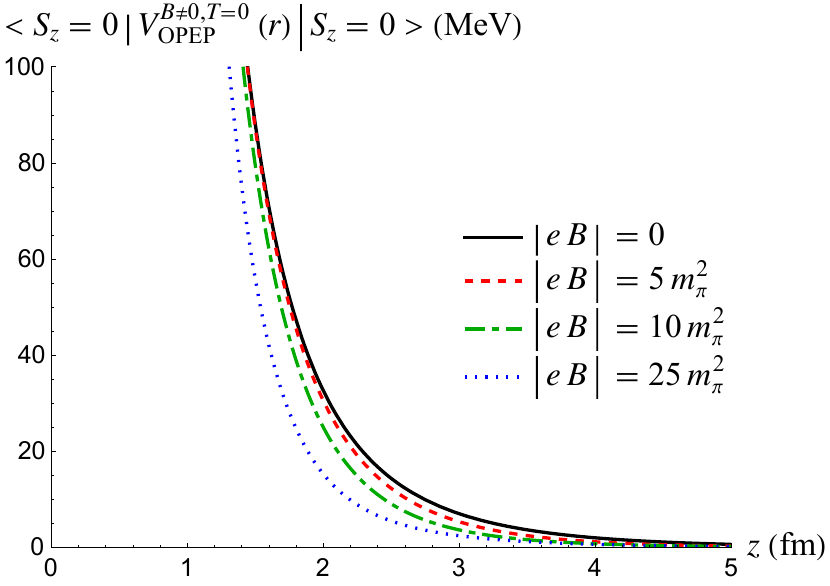}
  \end{minipage}
  \hfill
  \begin{minipage}{0.49\textwidth}
    \centering
    \adjustbox{raise=5.5mm}{\includegraphics[width=\textwidth]{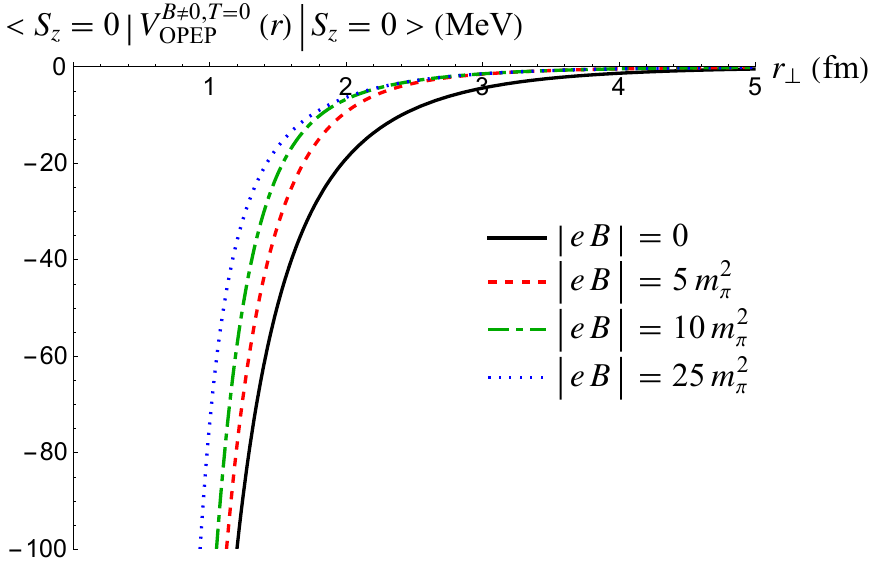}}
  \end{minipage}
  \caption{
    Magnetic-field dependence of the OPEP matrix element between the $\ket{S_z=0}$ states in the isospin-singlet ($T=0$) and spin-triplet ($S=1$) channel, as given in Eq.~\eqref{eq:potential-matrix-element-00}.
    Left panel: potential along the magnetic field (longitudinal) direction;
    right panel: potential along the perpendicular (transverse) direction.
    For the plots, we use the vacuum values: $g_A=1.27,\; f_\pi=92\text{ MeV}$, and $m_\pi=138\text{ MeV}$.
  }
  \label{fig:V00}
\end{figure*}

Figure~\ref{fig:V11} illustrates the behavior of the OPEP matrix element~\eqref{eq:potential-matrix-element-11} between the $S_z = +1$ states along both the $z$- and $\br_\perp$-directions for various values of the applied magnetic-field strength $|eB|$.
One can clearly observe that the range of the potential in both directions decreases as the magnetic-field strength increases.
This trend can be intuitively understood as a consequence of the Landau quantization of charged pions, which effectively increases their mass-squared in proportion to the field strength.

Having observed that the potential range decreases, however,
we cannot immediately conclude whether the net effect of the magnetic field enhances the attractive or repulsive components of the nucleon–nucleon interaction.
This is because the OPEP exhibits anisotropic behavior: it is attractive along the spin (magnetic-field) direction but repulsive in the transverse direction.
As a result, the shrinkage of the potential range in both directions can compete with each other, making it difficult to determine the overall net effect.

Figure~\ref{fig:V00} illustrates the behavior of the OPEP matrix element~\eqref{eq:potential-matrix-element-00} between the $S_z = 0$ states, plotted along the $z$- and $\br_\perp$-directions for various magnetic-field strengths $|eB|$.
Even without a magnetic field, the potential exhibits opposite attractive and repulsive behavior between the longitudinal and transverse directions, in contrast to $\braket{S_z=+1|V^{B\neq0,\,T=0}_\text{{OPEP}}(\br)|S_z=+1}$.
This directional dependence originates from the tensor operator $S_{12}$ in Eq.~\eqref{eq:tensor-operator}.
When a magnetic field is applied, the spatial rotational symmetry is explicitly broken, and the combination $\bm{\sigma}\cdot\br$, which is invariant under both spatial and spin rotations in the absence of the field, splits into $\bm{\sigma}_\perp\cdot\br_\perp$ and $\sigma^3z$.

\begin{figure*}[t]
  \centering
  \begin{minipage}{0.49\textwidth}
    \centering
    \includegraphics[width=\textwidth]{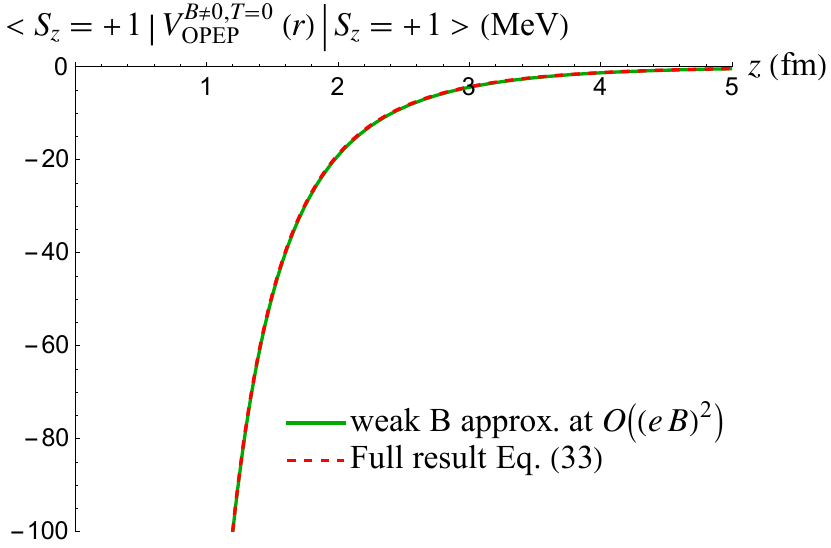}
  \end{minipage}
  \hfill
  \begin{minipage}{0.47\textwidth}
    \centering
    \includegraphics[width=\textwidth]{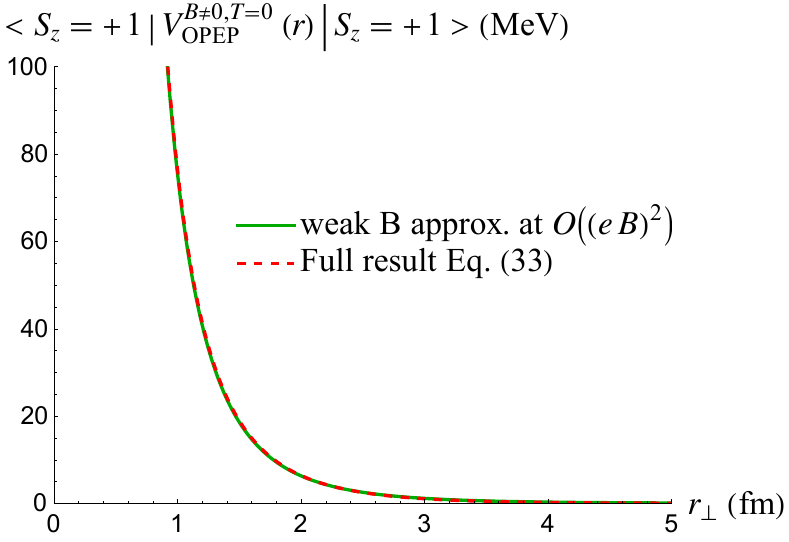}
  \end{minipage}
  \caption{
    Comparison of the exact result~\eqref{eq:potential-matrix-element-11} with the weak-magnetic-field expansion up to $\mathcal{O}\big( (eB)^2 \big)$ {derived from Eq.~(\ref{eq:OPEP-pn-weak})} for the OPEP matrix element between the $\ket{S_z=+1}$ states in the isospin-singlet ($T=0$) and spin-triplet ($S=1$) channel at $|eB|=m_\pi^2$.
  }
  \label{fig:V11_comparing-weak}
\end{figure*}

Let us also compare the full result with the weak-field and strong-field limits.
In Fig.~\ref{fig:V11_comparing-weak}, we compare the result of Eq.~\eqref{eq:potential-matrix-element-11} with that obtained from the weak-magnetic-field expansion up to $\mathcal{O}\big( (|eB|^2 \big)$.
One might expect that the weak-field expansion would break down around $|eB| \simeq m_\pi^2$.
However, due to the small numerical coefficients in the expansion [see the comment below Eq.~(\ref{eq:tensor-operator})], the weak-field approximation remains valid even near $|eB| \simeq m_\pi^2$.

\begin{figure*}[t]
  \centering
  \begin{minipage}{0.49\textwidth}
    \centering
    \includegraphics[width=\textwidth]{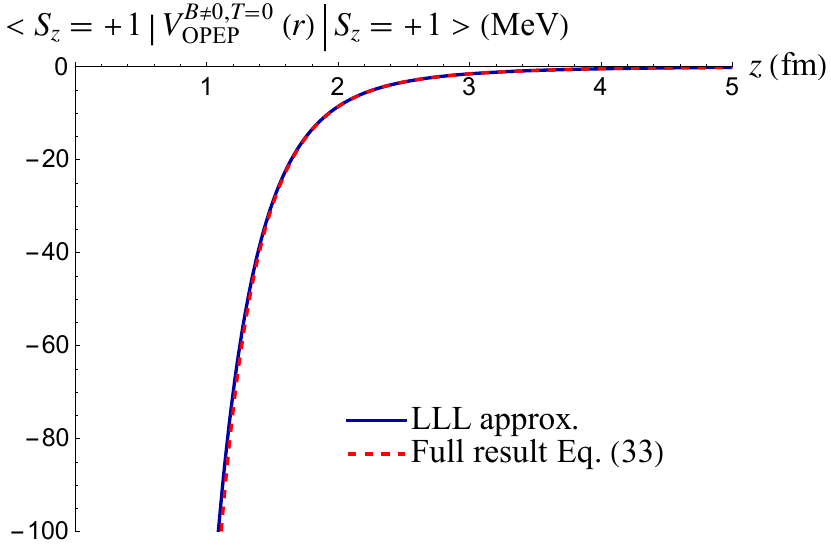}
  \end{minipage}
  \hfill
  \begin{minipage}{0.47\textwidth}
    \centering
    \includegraphics[width=\textwidth]{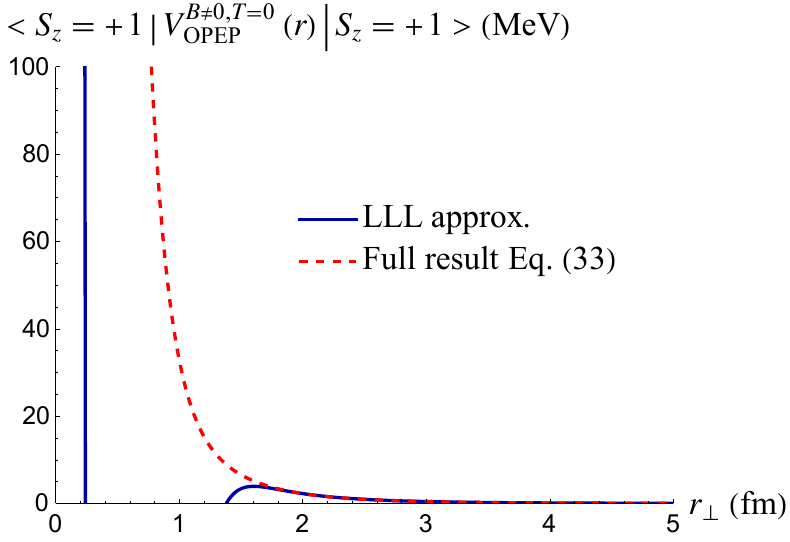}
  \end{minipage}
  \caption{
    Comparison of the exact result~\eqref{eq:potential-matrix-element-11}
    {with the strong-field LLL result based on Eq.~(\ref{eq:OPEP-pn-LLL})}
    for the OPEP matrix element between the $\ket{S_z=+1}$ states in the isospin-singlet ($T=0$) and spin-triplet ($S=1$) channel at $|eB|=25m_\pi^2$.
  }
  \label{fig:V11_comparing-LLL}
\end{figure*}

On the other hand, in Fig.~\ref{fig:V11_comparing-LLL}, we compare the full result with the lowest-Landau-level (LLL) approximation given in Eq.~\eqref{eq:OPEP-pn-LLL} at $|eB| = 25 m_\pi^2$.
As one can clearly see, the LLL contribution dominates in the long-distance region (e.g., $r_\perp \gtrsim 2$~fm in the plot).
This behavior is consistent with the condition for LLL dominance, $r_\perp \gg 1/\sqrt{|eB|}$.
At $|eB| = 25 m_\pi^2$, the magnetic length is $1/\sqrt{|eB|} = 1/(5m_\pi) \approx 0.3$~fm.
Therefore, at short transverse distances $r_\perp \lesssim 1/\sqrt{|eB|}$, higher-Landau-level contributions become non-negligible, and the LLL approximation deviates from the full result.

\subsection{OPEP in the isospin triplet channel}
\label{sec:OPEP-triplet}

We next examine the OPEP in the presence of a background magnetic field for the isospin-triplet ($T=1$) and spin-singlet ($S=0$) channel.
In what follows, we focus on the $T_z = 0$ component, where the charged-pion exchange contributes.
In contrast, for the $T_z = \pm 1$ components, only the neutral-pion exchange contributes, and thus the OPEP remains unmodified by the applied magnetic field.
As a result, we obtain the following matrix element:
\begin{align}
  \begin{split}
     & \braket{S_z=0|V^{B\neq0,\,T=1}_\text{{OPEP}}(\br)|S_z=0}                                                                                                                                                              \\
     & \hspace{10pt}
    = -V_0^{0,0}(\br)
    +\frac{g_A^2|eB|}{32\pi^2 f_\pi^2}\int_0^\infty \diff s\frac{1}{\sinh(|eB|s)}\sqrt{\frac{\pi}{s^3}}\left(1-\frac{1}{2s}z^2\right)\mathcal{F}(\br; s)
    \\
     & \hspace{30pt}+\frac{g_A^2|eB|^2}{32\pi^2f_\pi^2}\int_0^\infty \diff s\frac{1}{\sinh(|eB|s)}\frac{1}{\tanh(|eB|s)}\sqrt{\frac{\pi}{s}}\left(2-\frac{|eB|}{2}\frac{\br_\perp^2}{\tanh(|eB|s)}\right)\mathcal{F}(\br; s) \\
     & \hspace{40pt}+\frac{g_A^2|eB|^3}{64\pi^2f_\pi^2}\int_0^\infty \diff s\frac{\br_\perp^2}{\sinh(|eB|s)}\sqrt{\frac{\pi}{s}}\mathcal{F}(\br; s) ,
  \end{split}
  \label{eq:potential-matrix-element-00-isotriplet}
\end{align}

\begin{figure*}[t]
  \centering
  \begin{minipage}{0.49\textwidth}
    \centering
    \includegraphics[width=\textwidth]{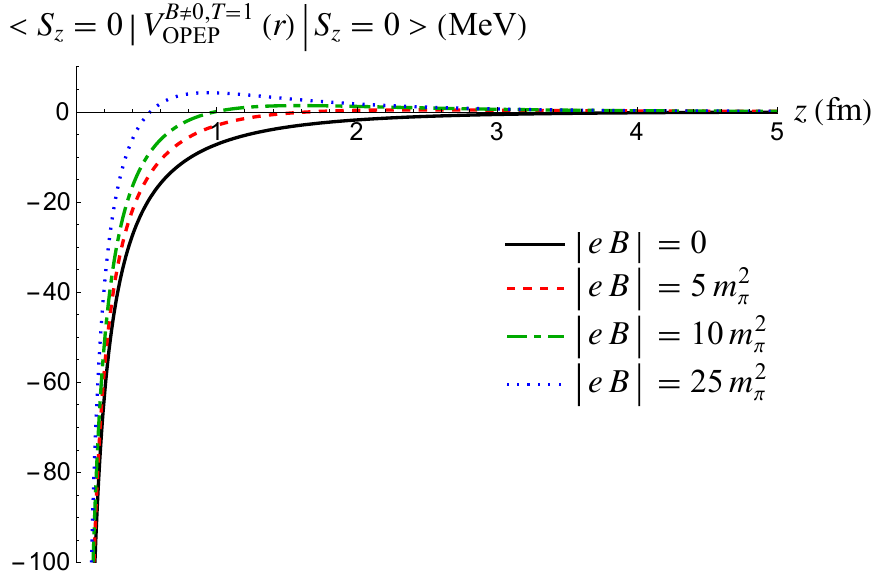}
  \end{minipage}
  \hfill
  \begin{minipage}{0.5\textwidth}
    \centering
    \includegraphics[width=\textwidth]{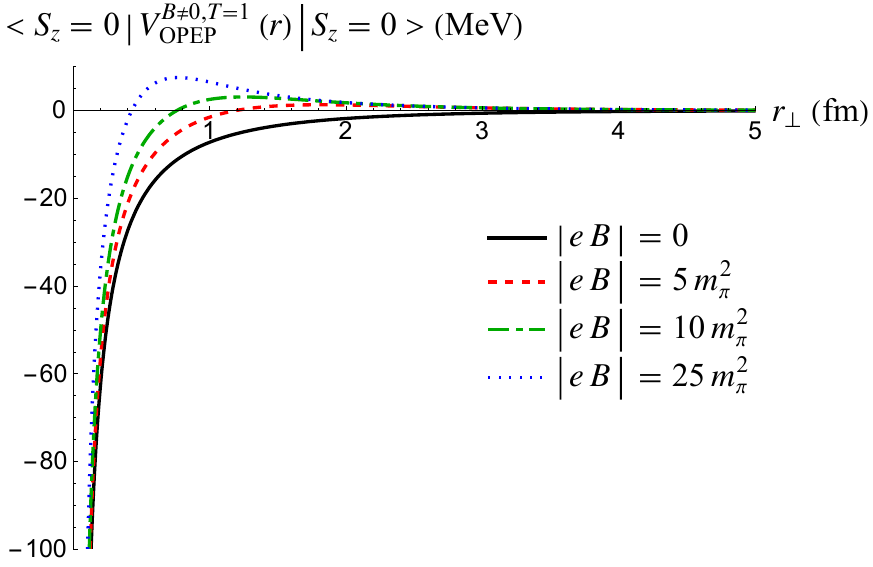}
  \end{minipage}
  \caption{
    Magnetic-field dependence of the OPEP matrix element between the $\ket{S_z=0}$ states in the isospin-triplet ($T=1$, especially $T_z=0$) and spin-singlet ($S=0$) channel, as given in Eq.~\eqref{eq:potential-matrix-element-00-isotriplet}.
    Left panel: potential along the magnetic field (longitudinal) direction;
    right panel: potential along the perpendicular (transverse) direction.
    For the plots, we use the vacuum values: $g_A=1.27, f_\pi=92\ \text{MeV}$, and $m_\pi=138\ \text{MeV}$.
  }
  \label{fig:V00isotriplet}
\end{figure*}
Figure~\ref{fig:V00isotriplet} shows the OPEP matrix element in the isospin-triplet channel with $T_z = 0$.
In the absence of a magnetic field, the OPEP reduces to a central, isotropic potential since the tensor operator $S_{12}$ in Eq.~\eqref{eq:tensor-operator} annihilates the spin-singlet state, $S_{12}\ket{S=0}=0$.
As a result, the potential is isotropic at $|eB|=0$.
Once a background magnetic field is introduced, however, one can see that the potential develops anisotropic features.

It is also interesting that the magnetic field induces a small bump in the potential at intermediate distances, causing the purely attractive potential at $|eB| = 0$ to develop a partially repulsive region.

\begin{figure*}[t]
  \centering
  \begin{minipage}{0.49\textwidth}
    \centering
    \includegraphics[width=\textwidth]{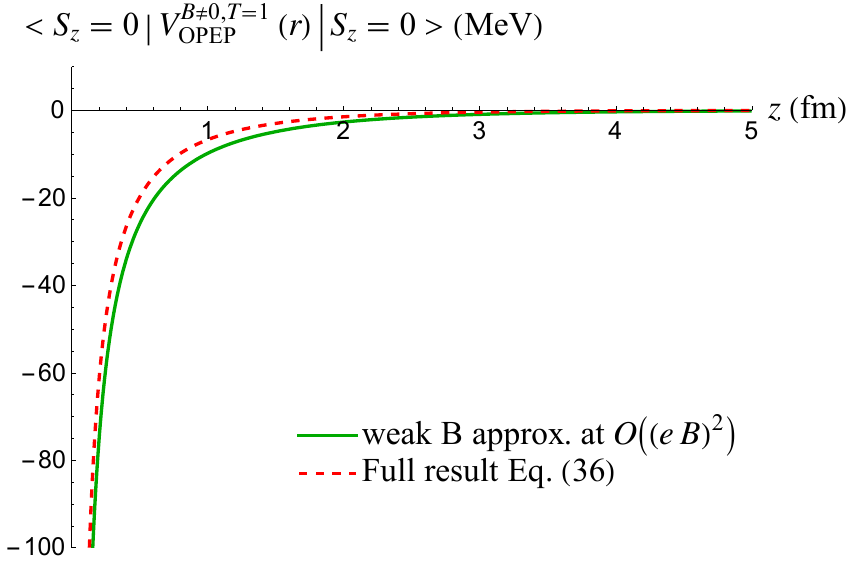}
  \end{minipage}
  \hfill
  \begin{minipage}{0.49\textwidth}
    \centering
    \includegraphics[width=\textwidth]{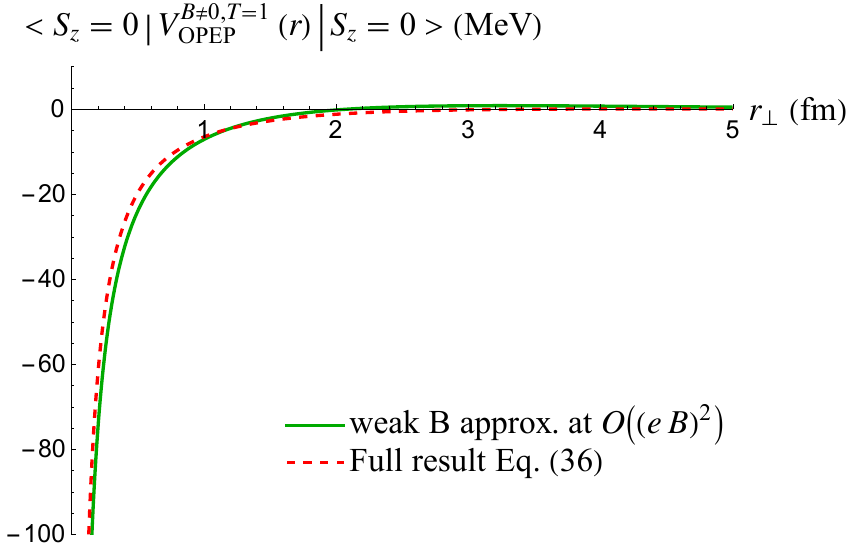}
  \end{minipage}
  \caption{
    Comparison of the exact result~\eqref{eq:potential-matrix-element-00-isotriplet} with the weak-magnetic-field expansion up to $\mathcal{O}\big( (eB)^2 \big)$ derived from Eq.~(\ref{eq:OPEP-pn-weak}) for the OPEP matrix element between the $\ket{S_z=0}$ states in the isospin-triplet ($T=1$) and spin-singlet ($S=0$) channel at $|eB|=m_\pi^2$.
  }
  \label{fig:Vtriplet_comparing-weak}
\end{figure*}

\begin{figure*}[t]
  \centering
  \begin{minipage}{0.49\textwidth}
    \centering
    \includegraphics[width=\textwidth]{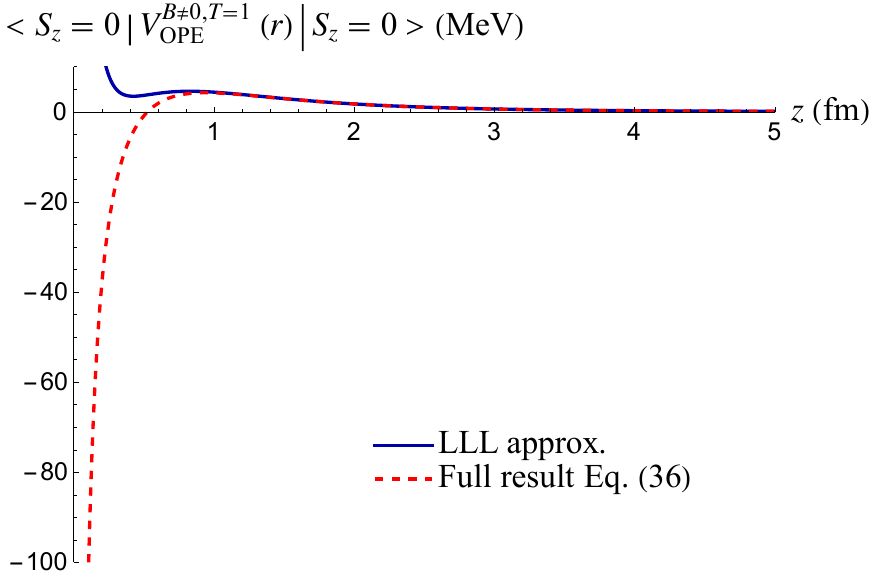}
  \end{minipage}
  \hfill
  \begin{minipage}{0.49\textwidth}
    \centering
    \includegraphics[width=\textwidth]{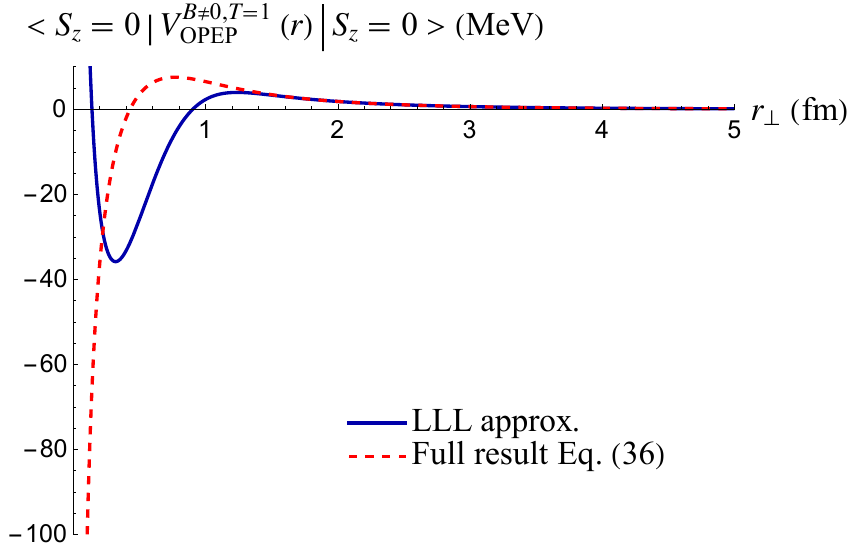}
  \end{minipage}
  \caption{
    Comparison of the exact result~\eqref{eq:potential-matrix-element-00-isotriplet}
    with the strong-field LLL result based on Eq.~(\ref{eq:OPEP-pn-LLL})
    for the OPEP matrix element between the $\ket{S_z=0}$ states in the isospin-triplet ($T=1$) and spin-singlet ($S=0$) channel at $|eB|=25m_\pi^2$.
  }
  \label{fig:Vtriplet_comparing-LLL}
\end{figure*}

We compare the full result with those of the weak-field and strong-field limits in Fig.~\ref{fig:Vtriplet_comparing-weak} and Fig.~\ref{fig:Vtriplet_comparing-LLL}.
The qualitative behavior is similar to that observed in the isospin-singlet channel.
In the weak-field regime, the perturbative expansion up to $\mathcal{O}((|eB|)^2)$ provides an excellent approximation to the full result even near $|eB|\sim m_\pi^2$.
In the strong-field regime, the lowest-Landau-level (LLL) approximation again reproduces only the long-distance behavior of the full result, while deviations appear at shorter separations due to the contributions from higher Landau levels.

\section{Energy shift of deuteron}
\label{sec:application}

In this section, we apply the derived OPEP to examine the impact of a magnetic field on the deuteron, which is the only two-nucleon bound state in the isospin-singlet ($T=0$) and spin-triplet ($S=1$).

We evaluate the energy shift of the deuteron in the presence of a magnetic field based on the derived OPEP, using the first-order perturbation theory.
Here, we focus solely on the modification of the nuclear force arising from the OPEP.
This assumption can be formally justified, since other magnetic-field-induced effects are further suppressed by a factor of $\mathcal{O}(|eB|/\Lambda_{\mathrm{UV}}^2)$, where $\Lambda_{\mathrm{UV}}$ is the ultraviolet cutoff of $\chi$EFT~(see also footnote~\ref{footnote2}).
Let $H_{\text{kin}}$ and $V_{\text{Heavy}}$ denote the kinetic term for the two-nucleon system and the potentials other than the OPEP, respectively, where $V_{\text{Heavy}}$ represents the short-range part of the nuclear force arising from heavier meson exchanges such as the $\rho$ and $\omega$ mesons.
In the absence of a magnetic field, the OPEP contains a tensor operator that mixes the $^3S_1$ and $^3D_1$ partial waves of the two-nucleon state.
As a consequence, the orbital angular momentum $L$ is not conserved, and the spin projection $S_z$ is mixed, while the total spin remains fixed at $S=1$.
The total angular momentum $J=L+S$, together with its projection $M=L_z+S_z$, thus serves as an appropriate quantum number to classify the deuteron states.
The eigenvalue problem for the deuteron can then be written as
\begin{align}
  (H_{\text{kin}} + V_{\text{Heavy}} + V_{\text{{OPEP}}}^{B=0}
  )
  \ket{d_{M}}
  = E^{B=0}\,\ket{d_{M}},
\end{align}
where $E^{B=0}\simeq -2.2\,\text{MeV}$ denotes the deuteron binding energy, and $M=0,\pm1$ labels the threefold degenerate eigenstates.

Now, we turn on a magnetic field.
For simplicity, we consider only the modification to the OPEP below.
In the one-boson-exchange picture, heavier charged mesons such as the $\rho^\pm$ also contribute to the nuclear force.
Since their magnetic-field dependence becomes relevant only at much higher field strengths, we restrict our analysis to the regime $|eB| \lesssim m_\pi^2$, where such effects may safely be neglected.
Accordingly, the corresponding short-range interaction $V_\text{Heavy}$ is treated as unaffected by the magnetic field.
We also neglect the direct coupling of the magnetic field to the nucleons themselves (e.g., the magnetic moments of the proton and neutron and the associated Zeeman splitting of the deuteron).
This means that possible spin polarization induced by the magnetic field is ignored.
These approximations justify focusing exclusively on the magnetic modification of the long-range part of the nuclear force—the OPEP. We note that the magnetic field couples not only to the charged-pion propagator but also to the $NN\pi$ coupling through the covariant derivatives in Eq.~(4).
Hence, the present $V_{\mathrm{{OPEP}}}^{B\neq0}$ defined in Eq.~(25) already contains the magnetic contribution to the vertex via the minimal coupling.

The problem is then reduced to solving an eigenvalue equation,
\begin{align}
  (H_\text{kin} + V_\text{Heavy} + V_\text{{OPEP}}^{B\neq 0})
  \ket{\tild_M^{B\neq0}}
  = E_{M}^{B\neq 0} \ket{\tild_M^{B \neq 0}},
\end{align}
where $E_M^{B\neq0}$ with $M=0,\pm1$ are the energy eigenvalues and $\ket{\tild_M^{B\neq0}}$ are the corresponding deuteron eigenstates in a magnetic field.

We then treat $V_\text{{OPEP}}^{B\neq0}-V_\text{{OPEP}}^{B=0}$ as a perturbation onto the unperturbed deuteron state without a magnetic field $\ket{d_M}$.
Using the degenerate perturbation theory, the first-order energy shift $\Delta E_{M}^{B\neq0}$ is determined by the eigenvalue equation,
\begin{align}
  A\bm{a}
   & = \Delta E_{M}^{B\neq0}\,\bm{a} ,
  \label{eq:eigen_deuteron}
\end{align}
where $\bm{a}$ is the coefficient vector that determines the linear combination of the unperturbed states, and the matrix $A = (A_{MM'})$ is defined as
\begin{align}
  A_{M M'}
   & :=
  \bra{d_M}
  V_\text{{OPEP}}^{B\neq0}-V_\text{{OPEP}}^{B=0}
  \ket{d_{M'}} .
\end{align}
For the unperturbed deuteron state (i.e., the wave function in the absence of the magnetic field), we use the wave function obtained from the AV18 potential, which includes both the ${}^3S_1$ and ${}^3D_1$ components.%
\footnote{We note that even if the unperturbed eigenstates in a magnetic field are defined as the dressed ones [i.e., with the Wilson lines~\eqref{eq:wilsoned-nucleon_field}], the matrix elements $A_{MM'}$ can still be evaluated using the undressed states, because the Wilson lines from the bra and ket cancel each other, i.e.,
  $
    \bra{\tilde d_M}
    \big(V_\text{OPEP}^{B\neq0}-V_\text{OPEP}^{B=0}\big)
    \ket{\tilde d_{M'}}
    =
    \bra{d_M}
    \big(V_\text{OPEP}^{B\neq0}-V_\text{OPEP}^{B=0}\big)
    \ket{d_{M'}} .
  $
}

\begin{figure}[t]
  \centering
  \includegraphics[width=0.5\linewidth]{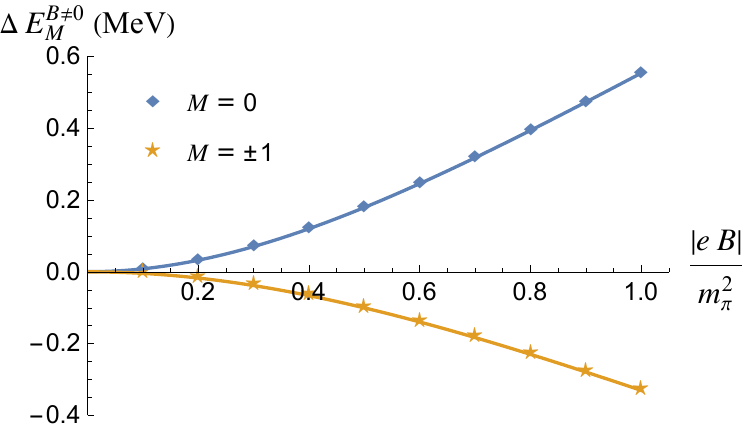}
  \caption{First-order energy shift of the deuteron $\Delta E^{B\neq 0}_{M}$ due to the OPEP in a strong magnetic field, plotted vs.~$|eB|$ up to $m_\pi^2$. The curves correspond to different degenerate states $M=0,\pm 1$.}
  \label{fig:energyshift}
\end{figure}

We numerically solve the eigenvalue equation~\eqref{eq:eigen_deuteron} and plot the resulting energy shift in Fig.~\ref{fig:energyshift}.
We observe that the magnetic-field-modified OPEP lifts the degeneracy between the $M =0$ and $M=\pm1$ states, while the degeneracy between $M =+1$ and $M=-1$ states remains intact.
The energy shifts for the $M \pm 1$ states are negative, indicating enhanced binding (i.e., increased stability), whereas the $M =0$  state receives a positive energy shift, implying reduced stability in a magnetic field.
We note that when the field becomes strong $|eB| \sim m_\pi^2$, the magnitude of the energy shift reaches $|\Delta E^{B \neq 0}_{M}| \sim 0.5\,$MeV, which is non-negligibly large compared to the binding energy in the absence of magnetic fields $E^{B=0} \simeq -2.2\,$MeV.

\section{Summary and discussion}
\label{sec:Summary}

In this paper, we have studied the modification to the one-pion exchange potential (OPEP) in a magnetic field.
Using chiral perturbation theory, we have derived the leading-order effective Lagrangian for pions and nonrelativistic nucleons, incorporating the magnetic-field effects on charged-pion propagation, as well as on the
$NN\pi$ vertex through the gauge potential in the covariant derivatives.
We have obtained the OPEP, which is made gauge invariant by redefining nucleon states by attaching a Wilson line.
The resulting OPEP involves a proper-time integration, which can also be expressed as a summation over the Landau levels.
We have evaluated the OPEP matrix elements for various channels such as the isospin-singlet and spin-triplet ($T=0, S=1$) and the isospin-triplet and spin-singlet ($T=1, S=0$) channels.
As a result, we have found that, in both channels, the range of the OPEP decreases as the magnetic-field strength increases.
The potential exhibits anisotropy in both channels, although its origin differs between them.
In the $T=0, S=1$ channel, as indicated by the weak-field limit~\eqref{eq:OPEP-pn-weak}, the anisotropy is dominated by the contribution from the tensor operator $S_{12}$, while the magnetic-field-induced anisotropy is suppressed by the small prefactor and thus is less effective.
In contrast, in the $T=1, S=0$ channel, where only the central force acts at zero magnetic field, a clear anisotropy appears once the magnetic field is applied.
Furthermore, we have applied the derived potential to compute the energy shift of the deuteron using the first-order perturbation theory.
We have found that the energy shift due to the modification of the OPEP by a magnetic field is of order 1\,MeV at $|eB|\sim m_\pi^2$.

Our results provide a first step toward understanding the impact of magnetic fields on nuclear force; or broadly speaking, the properties of nuclear force in external conditions.
Such a research direction is of interest by its own, helping us to develop a deeper understanding of nuclear force.

Let us discuss other possible extensions of our work.

First, additional magnetic-field effects that are not included in the present analysis should be taken into account.
For instance, the Zeeman energy shift is expected to be large when discussing the deuteron binding energy or nucleon bound states in general.
A rough estimate gives that the resulting energy shift is the order of $E_{\mathrm{Zeeman}} \sim eB/2m_N \sim 10\,\mathrm{MeV} $ at $|eB|\sim m_\pi^2$, which is much larger than the deuteron's binding energy $E^{B=0} \simeq -2.2\,$MeV.  Therefore, to obtain more realistic understanding of nuclear forces and their physical impacts in strong magnetic fields of order $|eB| \sim m_\pi^2$, these additional magnetic-field effects must be incorporated in future studies.

In addition, strong magnetic fields modify the masses of mesons mediating the nuclear force.
For example, the charged $\rho$-meson mass is expected to decrease with increasing magnetic field, and various theoretical studies also indicate that the neutral- and charged-pion masses change in strong magnetic fields (see, e.g., Refs.~\cite{Taya:2014nha,Bali:2017ian}).
Such modifications in the meson spectrum would feed back into the meson-exchange contributions to the nuclear force, including the OPEP.
Inclusion of these effects is an important direction toward a more quantitative description of nuclear forces in strong magnetic fields.

Another important theoretical challenge is to formulate a lattice QCD framework capable of computing nuclear forces in a background magnetic field---for example, by extending the HAL QCD method~\cite{Ishii:2006ec,Aoki:2009ji,Ishii:2012ssm}.

Finally, it is intriguing to explore the physical implications of our results in environments where strong magnetic fields are realized, such as the interiors of magnetars.
For example, the modification to OPEP may affect the modified URCA process, which changes the neutrino emissivity and in turn the cooling process of magnetars.
The equation-of-state of nuclear matter in magnetars can also be modified by the OPEP, whose impact may be discussed quantitatively based on the Brueckner theory, as was done in Ref.~\cite{Inoue:2013nfe} in the absence of magnetic fields.

\vspace{1em}
\section*{Acknowledgment}

The authors thank Koichi Hattori and Yoshimasa Hidaka for stimulating discussion.
D.M. is supported by JST SPRING, Grant No. JPMJSP2121 and JST, the establishment of University fellowships towards the creation of science technology innovation, Grant No. JPMJFS2114.
M.H. is supported by the Japan Society for the Promotion of Science (JSPS) KAKENHI Grants No. 23K25870, No. 25K01002, and No. 25K07316.
H.T. is supported by JSPS KAKENHI Grant No.~24K17058 and the RIKEN TRIP initiative (RIKEN Quantum).
T.H. was partly supported by Japan Science and Technology Agency (JST) as part of Adopting Sustainable Partnerships for Innovative Research Ecosystem (ASPIRE), Grant Number JPMJAP2318.
This work was partially supported by the RIKEN iTHEMS and Niigata University Quantum Research Center (NU-Q).

\appendix

\section{Charged pion Green's function in a strong magnetic field}
\label{sec:Green's function}

In this Appendix, we briefly review the derivation of Eq.~\eqref{eq:charge-pion-propag-space} in the main text, which gives the Green's function of charged pions (i.e., a charged scalar field) in a strong magnetic field.
We employ Schwinger's proper-time method~\cite{Schwinger:1951nm} to solve Eq.~\eqref{eq:propagator-spacerep}; see also the recent review~\cite{Hattori:2023egw} for further discussion.

As mentioned in the main text, we adopt the Fock-Schwinger symmetric gauge defined in Eq.~\eqref{eq:Fock-Schwinger-gauge}, and eventually recover the result in a general gauge by applying a gauge transformation.
The key observation in the Fock-Schwinger symmetric gauge is that the Klein-Gordon operator becomes translationally invariant.
Accordingly, we solve the equation of motion for the Green's function
\begin{align}
  \begin{split}
    \left[-
      \delta^{ij}
      \left(D^\pm_i\right)^\text{FS}\left(D^\pm_j\right)^\text{FS}
      + m_\pi^2
      \right]
    G_B({\bx-\bx'})
     & =\delta^{(3)}({\bx-\bx'}),
  \end{split}
  \label{eq:Klein-Gordon equation for Green's function in Fock-Schwinger gauge}
\end{align}
under Eq.~\eqref{eq:Fock-Schwinger-gauge}, which depends only on the relative coordinate {$\bx-\bx'$}.
As we will see shortly, the Green's function in this gauge turns out to be independent of the sign of the charge, so we omit the superscript $\pm$ hereafter.

We now consider the Green's function in four-dimensional momentum space and later take the static limit.
By performing a Fourier transform of the four-dimensional generalization of Eq.~\eqref{eq:Klein-Gordon equation for Green's function in Fock-Schwinger gauge}, we obtain
\begin{align}
  \begin{split}
    \left[
      p^2-m_\pi^2
      - \frac{\left(eB\right)^2}{4}\partial_{p_\perp}^2\
      + \rmi 0^+
      \right] G_B(p) = - 1,
  \end{split}
\end{align}
where $\rmi0^+$ (with $0^+$ denoting a positive infinitesimal) is introduced to ensure convergence of the proper-time integral,
and $p_\perp^\mu := (0,p_x, p_y,0)$ represents the transverse components of the momentum
(with respect to the magnetic field along the $z$ direction),
so that $\partial_{p_\perp}^2 := \partial^2/\partial p_x^2 + \partial^2/\partial p_y^2$.

This equation can be solved to yield the Green’s function expressed as a proper-time integral (see, e.g., Ref.~\cite{Hattori:2023egw}; note that our sign convention is opposite to that in the reference):
\begin{align}
  G_B(p) & =
  \rmi \int_0^\infty \diff s
  \frac{1}{\cos(eBs)}
  \exp\left(-\rmi(m_\pi^2-\rmi0^+)s+\rmi p_\parallel^2s-\rmi\frac{|\bp_\perp|^2
    }{eB}\tanh(eBs)\right)
  \nonumber                                                                                                        \\
         & =-2\rme^{-\frac{|\bp_\perp|^2}{|eB|}}\sum_{n=0}^\infty(-1)^nL_n\left(\frac{2|\bp_\perp|^2}{|eB|}\right)
  \frac{1}{p_\parallel^2-m_\pi^2-(2n+1)|eB|}.
  \label{eq:propagator-momentum}
\end{align}
Here, we introduced $p^2_\parallel :=p_0^2-p_z^2$
and used the $n$-th Laguerre polynomial $L_n (z)$.
The second line of this equation reveals the Landau-level structure, exhibiting an infinite number of discrete poles at $p_\parallel^2-m_\pi^2-(2n+1)|eB|=0$.
These poles correspond to the dispersion relations of charged pions undergoing Landau quantization in the magnetic field:
\begin{align}
  \begin{split}
    p^0=
    \pm
    \sqrt{p_z^2+m_\pi^2+(2n+1)|eB|} \with
    n = 0, 1,2,\cdots.
    \label{eq:specturm-Landau}
  \end{split}
\end{align}

Then, taking the static limit $p^0 \to 0$ and performing the Fourier transform, we obtain the spatial Green's function,
\begin{align}
  G_B( {\bx-\bx'}) & =\rmi\int \frac{\diff^3p}{(2\pi)^3}\rme^{-\rmi\bp\cdot({\bx-\bx'})}\int_0^\infty \diff s\frac{1}{\cos(eBs)}
  \exp\left(-\rmi(m_\pi^2-\rmi0^+)s-\rmi p_z^2s-\rmi\frac{|\bp_\perp|^2}{eB}\tanh(eBs)\right)
  \nonumber                                                                                                                      \\
                   & =\frac{|eB|}{8\pi^2}\int_0^\infty \diff
  s\frac{1}{\sin(|eB|s)}\sqrt{\frac{\pi}{\rmi s}}\exp
  \left(-\rmi(m_\pi^2-\rmi0^+)s+\rmi\frac{1}{4s}(z-z')^2+\rmi\frac{|eB|}{4}\frac{({\bx_\perp-\bx_\perp'})^2}{\tan(|eB|s)}\right).
  \label{eq:Green's function in Fock-Schwinger gauge}
\end{align}
It is worth noting that our Green's function~\eqref{eq:Green's function in Fock-Schwinger gauge} is related to the spatial Feynman propagator by time integral:
\begin{align}
  G_B({\bx-\bx'})=\int \diff t  G_B(t,{\bx-\bx'}).
\end{align}
One can also reexpress the proper-time integral with a Landau-level summation as
\begin{align}
  \begin{split}
    G_B({\bx-\bx'})
    = \frac{|eB|}{4\pi}\rme^{-\frac{|eB|}{4}|{\bx_\perp-\bx_\perp'}|^2}\sum_{n=0}^\infty L_n\left(\frac{|eB|}{2}|{\bx_\perp-\bx_\perp'}|^2\right)\frac{\rme^{-\sqrt{m_\pi^2+(2n+1)|eB|}|z-z'|}}{\sqrt{m_\pi^2+(2n+1)|eB|}}.
  \end{split}
  \label{eq:Green's function using Laguerre polynomial}
\end{align}
This result demonstrates that the screening mass along the $z$-direction is given by $\sqrt{m_\pi^2+(2n+1)|eB|}$ for each Landau level $n=0,1,2,\cdots$.

In numerical evaluations of the modified OPEP, the expression~\eqref{eq:Green's function using Laguerre polynomial} involving the Landau-level summation is not particularly useful due to its slow convergence up to moderate magnetic-field strengths.
We therefore employ a proper-time integral representation of the Green's function instead.
Moreover, we note that the present kinematics ($p^0=0$) allows us to deform the integration contour in Eq.~\eqref{eq:Green's function in Fock-Schwinger gauge}, which has poles on the real axis at $s=n\pi/|eB|$ with $n=0,1,2,\cdots$.
Since $m_\pi^2$ is positive and $0^+$ is a positive infinitesimal, we can deform the contour to avoid these poles into the lower-right region of the complex $s$-plane.
By Cauchy's integral theorem, we can then rewrite the expression as
\begin{align}
  G_B({\bx-\bx'})
   & =-\frac{|eB|}{8\pi^2}
  \int_{-\rmi\infty}^0 \diff s\frac{1}{\sin(|eB|s)}\sqrt{\frac{\pi}{\rmi s}}
  \exp
  \left(
  -\rmi(m_\pi^2-\rmi0^+)s+\rmi\frac{1}{4s}(z-z')^2+\rmi\frac{|eB|}{4}\frac{({\bx_\perp-\bx_\perp'})^2}{\tan(|eB|s)}
  \right)
  \nonumber                                      \\
   & =\frac{|eB|}{8\pi^2}\int_{0}^\infty \diff s
  \frac{1}{\sinh(|eB|s)}\sqrt{\frac{\pi}{s}}
  \exp\left(-m_\pi^2s-\frac{1}{4s}(z-z')^2-\frac{|eB|}{4}\frac{({\bx_\perp-\bx_\perp'})^2}{\tanh(|eB|s)}\right) .
  \label{eq:final-result-FS-gauge}
\end{align}
Here, the positive infinitesimal $0^+$ is dropped in the second line because the integral converges without it.
We use this representation instead of the infinite sum when plotting the potential and computing the deuteron energy shift.

Given the Green's function in the Fock–Schwinger symmetric gauge, Eq.~\eqref{eq:final-result-FS-gauge}, we can obtain its form in a general gauge by performing a gauge transformation:
\begin{align}
  \begin{split}
    A_i^\text{FS}\to A_i=A_i^\text{FS}-\partial_i\alpha.
  \end{split}
\end{align}
Recalling Eq.~\eqref{eq:Klein-Gordon equation for Green's function in Fock-Schwinger gauge}, this transformation leads to
\begin{align}
  \begin{split}
    G^\pm ({\bx,\bx'}|A)
     & = \rme^{\pm\rmi e\alpha({\bx}) \mp \rmi e\alpha({\bx'})}
    G_B({\bx-\bx'})
    = \rme^{\pm\rmi\Phi_A({\bx,\bx'})}
    G_B({\bx-\bx'}),                                            \\
    \Phi_A({\bx,\bx'})
     & = -\,e{\int_{\bm{x}'}^{\bm{x}}}
    \diff\bm{{\xi}}\cdot
    \Bigl[\bm{A}(\bm{{\xi}})+\tfrac12(\bm{{\xi}}-{\bm{x}'})\times\bm{B}\Bigr]\,,
  \end{split}
\end{align}
which leads to Eq.~\eqref{eq:charge-pion-propag-space} in the main text.
\clearpage

\section{Spin-dependent matrix elements in the isospin-singlet channel}
\label{sec:potential-matrix-element}

Here, we summarize all the $S_z$-dependent matrix elements for the potential from charged-pion exchange,
\begin{equation}
  \Delta V_\text{{OPEP}}^{B\neq0, T=0}(\br)
  {:=-V_{pn}(\br)-V_{np}(\br),}
\end{equation}
in the isospin-singlet ($T=0$) and spin-triplet ($S=1$) channel (excluding the neutral-pion exchange potential {$V_{0}$}
). Introducing the shorthand notation $\Delta V_\mathrm{{OPEP}}^{a,b} := \braket{S_z = a | \Delta V_\text{{OPEP}}^{B\neq0, T=0}(\br) | S_z = b}$, we obtain the following result:
\begin{align}
  \begin{split}
    \Delta V_{\mathrm{{OPEP}}}^{+1,+1}
     & =\frac{g_A^2|eB|}{32\pi^2 f_\pi^2}\int_0^\infty \diff s\frac{1}{\sinh(|eB|s)}\sqrt{\frac{\pi}{s^3}}
    \left(1-\frac{1}{2s}z^2\right)
    \mathcal{F}(\br; s),
    \\
    \Delta V_{\mathrm{{OPEP}}}^{+1,-1}
     & =-\frac{g_A^2|eB|^3}{64\pi^2 f_\pi^2}\int_0^\infty \diff s \frac{(x-\rmi y)^2}{\sinh(|eB|s)}\sqrt{\frac{\pi}{s}}\left(1+\frac{1}{\tanh^2(|eB|s)}\right) \mathcal{F}(\br; s),
    \\
    \Delta V_{\mathrm{{OPEP}}}^{+1,0}
     & =-\frac{\sqrt{2}g_A^2|eB|^2}{64\pi^2f_\pi^2}\int_0^\infty \diff s\frac{(x-\rmi y)z}{\sinh(|eB|s)}\frac{1}{\tanh(|eB|s)}\sqrt{\frac{\pi}{s^3}} \mathcal{F}(\br; s),
    \\
    \Delta V_{\mathrm{{OPEP}}}^{-1,+1}
     & =-\frac{g_A^2|eB|^3}{64\pi^2 f_\pi^2}\int_0^\infty \diff s \frac{(x+\rmi y)^2}{\sinh(|eB|s)}\sqrt{\frac{\pi}{s}}\left(1+\frac{1}{\tanh^2(|eB|s)}\right) \mathcal{F}(\br; s),
    \\
    \Delta V_{\mathrm{{OPEP}}}^{-1,-1}
     & =\frac{g_A^2|eB|}{32\pi^2 f_\pi^2}\int_0^\infty \diff s\frac{1}{\sinh(|eB|s)}\sqrt{\frac{\pi}{s^3}}\left(1-\frac{1}{2s}z^2\right) \mathcal{F}(\br; s),
    \\
    \Delta V_{\mathrm{{OPEP}}}^{-1,0}
     & =-\frac{\sqrt{2}g_A^2|eB|^2}{64\pi^2f_\pi^2}\int_0^\infty \diff s\frac{(x+\rmi y)z}{\sinh(|eB|s)}\frac{1}{\tanh(|eB|s)}\sqrt{\frac{\pi}{s^3}} \mathcal{F}(\br; s),
    \\
    \Delta V_{\mathrm{{OPEP}}}^{0,+1}
     & =-\frac{\sqrt{2}g_A^2|eB|^2}{64\pi^2f_\pi^2}\int_0^\infty \diff s\frac{(x+\rmi y)z}{\sinh(|eB|s)}\frac{1}{\tanh(|eB|s)}\sqrt{\frac{\pi}{s^3}} \mathcal{F}(\br; s),
    \\
    \Delta V_{\mathrm{{OPEP}}}^{0,-1}
     & =-\frac{\sqrt{2}g_A^2|eB|^2}{64\pi^2f_\pi^2}\int_0^\infty \diff s\frac{(x-\rmi y)z}{\sinh(|eB|s)}\frac{1}{\tanh(|eB|s)}\sqrt{\frac{\pi}{s^3}} \mathcal{F}(\br; s),
    \\
    \Delta V_{\mathrm{{OPEP}}}^{0,0}
     & =-\frac{g_A^2|eB|}{32\pi^2 f_\pi^2}\int_0^\infty \diff s\frac{1}{\sinh(|eB|s)}\sqrt{\frac{\pi}{s^3}}\left(1-\frac{1}{2s}z^2\right) \mathcal{F}(\br; s),
    \\
     & \hspace{12pt}+\frac{g_A^2|eB|^2}{32\pi^2f_\pi^2}\int_0^\infty \diff s\frac{1}{\sinh(|eB|s)}\frac{1}{\tanh(|eB|s)}\sqrt{\frac{\pi}{s}}\left(2-\frac{|eB|}{2}\frac{\br_\perp^2}{\tanh(|eB|s)}\right) \mathcal{F}(\br; s) \\
     & \hspace{12pt}+\frac{g_A^2|eB|^3}{64\pi^2f_\pi^2}\int_0^\infty \diff s\frac{\br_\perp^2}{\sinh(|eB|s)}\sqrt{\frac{\pi}{s}} \mathcal{F}(\br; s),
  \end{split}
  \label{matrix-element-spin}
\end{align}
where the exponential factor $\mathcal{F}(\br; s)$ has been defined in Eq.~(\ref{eq:def-F}) in the main text.

\bibliographystyle{utphys}
\bibliography{nulcear-force-B}

\end{document}